\documentclass[preprint,showpacs,preprintnumbers,amsmath,amssymb]{revtex4}

\usepackage{graphicx}
\usepackage{dcolumn}
\usepackage{bm}

\begin{document}

\preprint{}

\title{Dyson Indices and 
Hilbert-Schmidt Separability Functions and Probabilities}

\author{Paul B. Slater}%
\email{slater@kitp.ucsb.edu}
\affiliation{%
ISBER, University of California, Santa Barbara, CA 93106\\
}%
\date{\today}

\begin{abstract}
A confluence of numerical and theoretical results
leads us to conjecture that the Hilbert-Schmidt separability
{\it probabilities}
of the 15- and 9-dimensional convex sets of complex and real two-qubit states 
(representable by $4 \times 4$ density matrices $\rho$) are 
$\frac{8}{33}$ and $\frac{8}{17}$, respectively. Central to our
reasoning are the modifications of 
two ans{\"a}tze, recently advanced ({\it Phys. Rev. A}, {{\bf{75}}} 
[2007], 032326), involving incomplete beta functions 
$B_{\nu}(a,b)$, where $\nu= 
\frac{\rho_{11} \rho_{44}}{\rho_{22} \rho_{33}}$. We, now, set the 
separability {\it function} $\mathcal{S}_{real}(\nu)$ proportional to $
B_{\nu}(\nu,\frac{1}{2},2)
=\frac{2}{3} (3-\nu) \sqrt{\nu}$.
Then, in the {\it complex} case---conforming to a pattern we find,  
manifesting
the Dyson indices ($\beta=1, 2, 4$) 
of random matrix theory--we take $\mathcal{S}_{complex}(\nu)$ proportional to
$\mathcal{S}_{real}^{2}(\nu)$.
We also investigate the real and complex qubit-{\it qutrit} cases.
Now, there are {\it two} variables, 
$\nu_{1}= \frac{\rho_{11} \rho_{55}}{\rho_{22} \rho_{44}},
\nu_{2}= \frac{\rho_{22} \rho_{66}}{\rho_{33} \rho_{55}}$,
but they appear to remarkably coalesce into the product $\eta = \nu_1 \nu_2 = 
 \frac{\rho_{11} \rho_{66}}{\rho_{33} \rho_{44}}$, so that
the real and complex separability functions are 
again {\it univariate} in nature.
\end{abstract}


\maketitle
\tableofcontents
\section{Introduction}
\.Zyczkowski and Sommers
have derived---using random matrix theory (in particular, the Laguerre 
ensemble)---general 
formulas for the $(n^2-1)$-dimensional 
and the \newline $\frac{n (n+1)-2}{2}$-dimensional 
volumes of the complex and real 
$n \times n$ density matrices ($\rho$), 
respectively,  in terms of the Hilbert-Schmidt (HS) 
metric  \cite{szHS} \cite[sec. 14.3]{ingemarkarol} 
(as well as the Bures metric 
\cite[sec. 14.4]{ingemarkarol} 
\cite{szBures,slaterJGP}). Later, Andai \cite{andai} 
examined these and related questions, using a quite different 
framework. He applied mathematical induction on
the leading principal minors of $\rho$, along with the established 
formulas for 
hyperareas of surfaces of $n$-spheres and beta integrals. He   
reproduced---up to normalization factors---the HS real and complex 
volume 
formulas in \cite{szHS} (and, moreover,
the $(2 n^2 -n -1)$-dimensional quaternionic volumes). 
(In addition to the HS and Bures metrics, 
Andai considered, for the single qubit case, 
the broad [infinitely nondenumerable] 
class---which does include the Bures as
its minimal member---of 
monotone metrics. Unlike {\.Z}yczkowski and Sommers, he did not obtain
formulas for the 
hyperareas occupied by density matrices of {\it less} than full rank.)

Despite these considerable theoretical advances, 
volume (and, hence, probability) 
formulas have not yet become available for the 
important subsets of separable
($n \leq 6$) and positive-partial-transpose ($n \geq 8$) 
$n \times n$ density matrices ($n$ composite).
(Szarek \cite{sz1}  employed methods
of {\it asymptotic} convex geometry to {\it estimate} the volume of
the set of separable mixed quantum states for $N$ qubits, and Aubrun and Szarek
\cite{sz2} for $N$ qudits. It was concluded in these studies 
that the separable volumes 
were superexponentially small in the dimension of the set of states. 
For large $D$, the $(D^2-1)$-dimensional volume for bipartite systems 
of positive-partial-transpose 
states, however, is much larger than the volume of separable states 
\cite[Thm. 4]{sz2}.)

To address this fundamental lacuna, at least in the Hilbert-Schmidt 
context (cf. \cite{BuresSep}), we 
developed in \cite{slaterPRA2} a methodology---incorporating the Bloore
parameterization of density matrices \cite{bloore} (sec.~\ref{fjparam}). 
Its numerical application
led to ans{\"a}tze, involving (apparently independent) 
incomplete beta functions, 
for the 9-dimensional real and 15-dimensional 
complex separable volumes in the qubit-qubit ($n=4$) 
case \cite{slaterPRA2}. In the sequel to that study here, 
we, first, apply this Bloore framework to various scenarios involving 
$n \times n$ density matrices 
($n =4, 6, 8, 9$), in which certain of 
their off-diagonal entries have been {\it nullified}.
This enables us  to now obtain {\it exact} results, of interest in 
themselves, and 
possibly suggestive of  solutions/approaches 
to the full (non-nullified) highly 
computationally-challenging problems.

In fact, based on certain 
(real-complex-quaternionic) 
patterns emerging in these exact results 
(sec.~\ref{DysonIndex}), bearing
an obvious relation to the Dyson indices ($\beta=1, 2, 4$) of random matrix theory 
\cite{dyson}, we are 
led to modify the incomplete beta function ans{\"a}tze for the two full 
(real and complex)
problems 
advanced in \cite{slaterPRA2}.
The ``separability function'' in the complex case is now
{\it not} analyzed as if it were independent of that
in the real case  (which we still take to be
an incomplete---but slightly different---beta function), 
but actually simply proportional to its {\it square}. These 
central analyses will be elaborated upon in sec.~\ref{twoqubitconjectures}, 
(eqs. (\ref{almost1}) - (\ref{almost3})), where it shown that the modified
ans{\"a}tze do, in fact, accord well (Fig.~\ref{fig:QubQut}) 
with the numerical results of \cite{slaterPRA2}.

We begin  our extensive series of lower-dimensional analyses, by
examining a number of two-qubit scenarios (sec.~\ref{qubit-qubit}).
In them, we are able to compute 
a number of interesting
exact two-qubit scenario-specific
HS separability probabilities. (Listing them in increasing order, we have 
$\left\{\frac{1}{10}, 
\frac{1}{3},\frac{3}{8},\frac{2}{5},\frac{135 \pi
   }{1024},\frac{16}{3 \pi ^2},\frac{3 \pi
   }{16},\frac{5}{8},\frac{105 \pi }{512},2-\frac{435 \pi
   }{1024},\frac{11}{16},1\right\}$.)
For each of the scenarios, we 
identify a certain univariate 
separability function $\mathcal{S}_{scenario}(\nu)$, 
where $\nu = \frac{\rho_{11} \rho_{44}}{\rho_{22} \rho_{33}}$. 
The integral over $\nu \in [0,\infty]$ of the product of this
function (typically of a {\it piecewise} nature over [0,1] and $[1,\infty]$) 
with a scenario-specific (marginal) jacobian function 
$\mathcal{J}_{scenario}(\nu)$ yields
the HS {\it separable} volume ($V^{HS}_{sep}$). The ratio of 
$V^{HS}_{sep}$ to the
HS {\it total} (entangled and non-entangled) 
volume ($V^{HS}_{tot}$) 
gives us the HS scenario-specific separability {\it probability}.

The question of the 
``relative proportion'' of entangled and non-entangled states 
in a given generic class of composite
quantum systems,  had
apparently first been raised by {\. Z}yczkowski, Horodecki, 
Sanpera and Lewenstein (ZHSL) in a much-cited paper \cite{ZHSL}.
They gave  ``three main reasons''---``philosophical'', ``practical'' 
and ``physical''---upon which they expanded, for pursuing the topic.
The present author, motivated by the ZHSL paper,  
has investigated this issue in a number of  settings, using 
various (monotone and non-monotone) measures on  
quantum states, and a variety of
numerical and analytical methods
\cite{slaterC,slaterA,slaterOptics,slaterqip,
pbsJak,slaterPRA,slaterJGP,pbsCanosa} 
(cf. \cite{sbz,sepsize1,sepsize2,sepsize3}).
Though the problems are challenging 
(high-dimensional) in nature, many of the results obtained 
in answer to the ZHSL question in these various contexts 
have been strikingly simple and elegant 
(and/or conjecturally so).

Specifically here, we further develop the 
(Bloore-parameterization-based) approach presented in
\cite{slaterPRA2}. This was found to be relatively effective in
studying the question posed by ZHSL, in the context of two-qubit
systems (the smallest possible example exhibiting entanglement), 
endowed with the
(non-monotone \cite{ozawa}) 
{\it Hilbert-Schmidt}  (HS) measure \cite{szHS}, inducing the flat,
{\it Euclidean} geometry on the space of $4 \times 4$ density matrices.
This approach \cite{slaterPRA2} exploits two distinct features of a
form of density matrix parameterization first discussed by Bloore 
\cite{bloore}. 
These properties  allow 
us to deal with lower-dimensional integrations
(more amenable to computation) than would otherwise be possible.
We further find that the interesting advantages 
of the Bloore parameterization do, in fact,  carry over---in a 
somewhat modified fashion---to the qubit-qutrit 
(sec.~\ref{qubitqutrit} and \ref{qubqut2}), qutrit-qutrit (sec.~\ref{qutritqutrit}) 
and qubit-qubit-qubit (secs.~\ref{qubitqubitqubitI} and 
\ref{qubitqubitqubitII}) domains.
\subsection{Bloore (off-diagonal-scaling) parameterization} \label{fjparam}
We, first, consider the 9-dimensional convex set of 
(two-qubit) $4 \times 4$ density matrices with {\it real} entries,
and parameterize them---following Bloore \cite{bloore} 
(cf. \cite[p. 235]{kurowicka})---as 
\begin{equation} \label{BlooreDenMat}
\rho = \left(
\begin{array}{llll}
 \rho _{11} & z_{12} \sqrt{\rho _{11} \rho _{22}} & z_{13}
   \sqrt{\rho _{11} \rho _{33}} & z_{14} \sqrt{\rho _{11}
   \rho _{44}} \\
 z_{12} \sqrt{\rho _{11} \rho _{22}} & \rho _{22} & z_{23}
   \sqrt{\rho _{22} \rho _{33}} & z_{24} \sqrt{\rho _{22}
   \rho _{44}} \\
 z_{13} \sqrt{\rho _{11} \rho _{33}} & z_{23} \sqrt{\rho
   _{22} \rho _{33}} & \rho _{33} & z_{34} \sqrt{\rho
   _{33} \rho _{44}} \\
 z_{14} \sqrt{\rho _{11} \rho _{44}} & z_{24} \sqrt{\rho
   _{22} \rho _{44}} & z_{34} \sqrt{\rho _{33} \rho _{44}}
   & \rho _{44}
\end{array}
\right).
\end{equation}
One, of course, has the 
standard requirements that $\rho_{ii} \geq 0$ and 
(the unit trace condition) $\Sigma_{i} \rho_{ii} 
= 1$.
Now, three additional necessary conditions
(which can be expressed {\it without} using the diagonal entries,  
due to the  $\rho_{ii} \geq 0$ stipulation) 
that must be fulfilled 
for $\rho$ to be a
density matrix (with all eigenvalues non-negative) are: (1) 
the non-negativity of the determinant (the principal $4 \times 4$ minor),
\begin{equation} \label{firstcondition}
\left(z_{34}^2-1\right) z_{12}^2+2 \left(z_{14}
   \left(z_{24}-z_{23} z_{34}\right)+z_{13}
   \left(z_{23}-z_{24} z_{34}\right)\right)
   z_{12} -z_{23}^2-z_{24}^2-z_{34}^2 +
\end{equation}
\begin{displaymath}
+ z_{14}^2    \left(z_{23}^2-1\right)+z_{13}^2
   \left(z_{24}^2-1\right)+2 z_{23} z_{24} z_{34}+2 z_{13}
   z_{14} \left(z_{34}-z_{23} z_{24}\right)+1 \geq 0;
\end{displaymath}
(2): 
the non-negativity of the leading principal $3 \times 3$ minor,
\begin{equation} \label{secondcondition}
-z_{12}^2+2 z_{13} z_{23} z_{12}-z_{13}^2-z_{23}^2+1 \geq 0;
\end{equation}
and (3): the non-negativity of the principal $2 \times 2$ minors 
(although actually only the $i=j=1$ case is needed, it is natural
to impose them all),
\begin{equation} \label{2by2}
1-z_{ij}^2 \geq 0.
\end{equation}
As noted,
the {\it diagonal} entries of $\rho$ do {\it not} enter into any of 
these constraints---which taken together are {\it sufficient} to guarantee
the nonnegativity of $\rho$ itself---as they can be shown to 
contribute only (cancellable) non-negative factors to the
determinant and principal minors. This cancellation property 
is certainly 
a principal virtue of the Bloore parameterization, allowing one to
proceed analytically in lower dimensions than one might 
initially surmise.
(Let us note that, utilizing this parameterization,
we have been able to establish a recent 
conjecture of M{\aa}nsson, Porta Mana and Bj{\"o}rk regarding Bayesian state 
assignment for three-level quantum systems, and, in fact, verify our own
four-level analogue of their conjecture \cite[eq. (52)]{mansson} 
\cite{mansson2}.)

Additionally, implementing the Peres-Horodecki condition
\cite{asher,michal,bruss} requiring
the non-negativity of the {\it partial transposition} of $\rho$,
we have the 
necessary {\it and} sufficient 
condition for the {\it separability} (non-entanglement) of  $\rho$ 
that (4):
\begin{equation} \label{Thirdcontion}
\nu  \left(z_{34}^2-1\right) z_{12}^2+2 \sqrt{\nu }
   \left(\nu  z_{13} z_{14}+z_{23} z_{24}-\sqrt{\nu }
   \left(z_{14} z_{23}+z_{13} z_{24}\right) z_{34}\right)
   z_{12}-z_{23}^2-\nu  z_{34}^2+\nu +
\end{equation}
\begin{displaymath}
 +\nu 
   \left(\left(z_{24}^2-1\right) z_{13}^2-2 z_{14} z_{23}
   z_{24} z_{13}-z_{24}^2+z_{14}^2 \left(z_{23}^2-\nu
   \right)\right)+2 \sqrt{\nu } \left(z_{13} z_{23}+\nu 
   z_{14} z_{24}\right) z_{34} \geq 0,
\end{displaymath}
where
\begin{equation} \label{BlooreRatio}
\nu = \mu^2 = \frac{\rho_{11} \rho_{44}}{\rho_{22} \rho_{33}},
\end{equation}
being the only information needed, at this stage, concerning
the diagonal entries of $\rho$. (It is interesting to contrast
the role of our variable $\nu$, as it pertains to the determination of
entanglement, with the rather different roles played by 
the {\it concurrence} and {\it negativity} \cite{ver2,imai}.)
We have vacillated between the use of $\nu$ and $\mu$ as our principal
variable in our two previous studies \cite{univariate,slaterPRA2}. In 
sec.~\ref{approximate}, we will revert to the use of $\mu$, as it appears that
its use can avoid the appearances of square roots, which, it is our 
impression, at least, can impede certain Mathematica computations.
\subsubsection{Reduction of dimensionality}
Thus, the Bloore parameterization is 
evidently even further convenient here, in reducing
the apparent dimensionality of the separable volume problem. That is, we 
now have to essentially consider only the separability variable 
$\nu$ rather than {\it three} independent (variable)
diagonal entries. 
(This supplementary feature had not been commented upon 
by Bloore, as he discussed only
$2 \times 2$ and $3 \times 3$ density matrices, and also, obviously,
since the Peres-Horodecki 
separability condition had not yet been formulated in 1976.)
The ({\it two} variable---$\nu_{1}, \nu_{2}$) 
analogue of (\ref{BlooreRatio}) in the $6 \times 6$ 
(qubit-{\it qutrit}) case will be discussed 
and implemented in sec.~\ref{qubitqutrit}.
Additionally still, we find a {\it four}-variable counterpart in
the $9 \times 9$ qutrit-qutrit instance [sec.~\ref{qutritqutrit}], 
and {\it three}-variable counterparts in two sets of qubit-qubit-qubit analyses
(secs.~\ref{qubitqubitqubitI} and \ref{qubitqubitqubitII}).
(The question of whether any or all of these several 
ratio variables are
themselves {\it observables} would seem to be of some interest.) 
It certainly appears to us that in the qubit-qutrit case
(sec.~\ref{qubitqutrit} and \ref{qubqut2}) the two associated
ratio variables ($\nu_1, \nu_2$) 
importantly merge or coalesce into the simple product
$\eta = \nu_1 \nu_2$ for all analytical purposes. 
({\it Products} of ratio variable do also appear in the limited number
of still higher-dimensional analyses we report below, so perhaps some similar
merging or coalescing takes place in those settings, as well.)

In \cite[eqs. (3)-(5)]{slaterPRA2}, we expressed the conditions
(found through application of the ``cylindrical algebraic decomposition'' 
\cite{cylindrical})
that---in terms of the Bloore variables $z_{ij}$'s---an 
arbitrary  9-dimensional $4 \times 4$ real density matrix 
$\rho$ must fulfill.
These took the form,
\begin{equation} \label{limits}
 z_{12}, z_{13}, z_{14} \in [-1,1],
 z_{23} \in [Z^-_{23},Z^+_{23}],
 z_{24} \in [Z^-_{24},Z^+_{24}],
 z_{34} \in [Z^-_{34},Z^+_{34}],
\end{equation}
where
\begin{equation}
Z^{\pm}_{23} =z_{12} z_{13} \pm \sqrt{1-z_{12}^2} \sqrt{1-z_{13}^2} , 
Z^{\pm}_{24} =z_{12} z_{14} \pm \sqrt{1-z_{12}^2} \sqrt{1-z_{14}^2} ,
\end{equation}
\begin{displaymath}
Z^{\pm}_{34} = \frac{z_{13} z_{14} -z_{12} z_{14} z_{23} -z_{12} z_{13} z_{24} +z_{23}
z_{24} \pm s}{1-z_{12}^2},
\end{displaymath}
and
\begin{equation} \label{limits2}
s = \sqrt{-1 +z_{12}^2 +z_{13}^2 -2 z_{12} z_{13} z_{23} +z_{23}^2}
\sqrt{-1 +z_{12}^2 +z_{14}^2 -2 z_{12} z_{14} z_{24} +z_{24}^2}.
\end{equation}
\subsubsection{Possible transformations of the $z_{ij}$'s}
In his noteworthy paper, Bloore also presented \cite[secs. 6,7]{bloore} 
a quite interesting
discussion of the ``spheroidal'' geometry induced by his parameterization.
This strongly suggests that it might prove useful to reparameterize 
the $z_{ij}$ variables in terms of spheroidal-type coordinates.
Following the argument of Bloore---that is, performing rotations
of the $(z_{13},z_{23})$ and $(z_{14},z_{24})$ 
vectors by $\frac{\pi}{4}$ and recognizing that each pair of so-transformed
variables lay in ellipses with axes of length 
$\sqrt{1 \pm z_{12}}$---we were able to 
substantially simply
the forms of the feasibility  conditions ((\ref{limits})-(\ref{limits2})).

Using the set of transformations (having a jacobian equal to 
$\frac{\left(1-z_{12}^2\right) \gamma _1}{2 \gamma _2}$)
\begin{equation} \label{reparameterization}
z_{13}\to \frac{\left(\sqrt{1-z_{12}} \cos
   \left(\theta _1\right)+\sin \left(\theta _1\right)
   \sqrt{z_{12}+1}\right) \sqrt{\gamma _1 \gamma
   _2+1}}{\sqrt{2}},
\end{equation}
\begin{displaymath}
z_{23}\to \frac{\left(\sin
   \left(\theta _1\right) \sqrt{z_{12}+1}-\cos
   \left(\theta _1\right) \sqrt{1-z_{12}}\right)
   \sqrt{\gamma _1 \gamma _2+1}}{\sqrt{2}},
\end{displaymath}
\begin{displaymath}
z_{14}\to
   \frac{\left(\sqrt{1-z_{12}} \cos \left(\theta
   _2\right)+\sin \left(\theta _2\right)
   \sqrt{z_{12}+1}\right) \sqrt{\gamma _1+\gamma
   _2}}{\sqrt{2} \sqrt{\gamma _2}},
\end{displaymath}
\begin{displaymath}
z_{24}\to
   \frac{\left(\sin \left(\theta _2\right)
   \sqrt{z_{12}+1}-\cos \left(\theta _2\right)
   \sqrt{1-z_{12}}\right) \sqrt{\gamma _1+\gamma
   _2}}{\sqrt{2} \sqrt{\gamma _2}},
\end{displaymath}
\begin{displaymath}
z_{34}\to
   Z_{34}-\frac{\cos \left(\theta _1-\theta _2\right)
   \sqrt{\gamma _1+\gamma _2} \sqrt{\gamma _1 \gamma
   _2+1}}{\sqrt{\gamma _2}} ,
\end{displaymath}
one is able to replace the conditions ((\ref{limits})-(\ref{limits2})) 
that the real two-qubit 
density matrix $\rho$---given by (\ref{BlooreDenMat})---must 
fulfill 
by
\begin{equation} \label{newlimits}
\gamma_{1} \in [0,1]; \hspace{.1in} \gamma_{2} \in [\gamma_{1},\frac{1}{\gamma_{1}}]; \hspace{.1in}
Z_{34} \in [-\gamma_{1}, \gamma_{1}]; \hspace{.1in} z_{12} \in [-1,1]; \hspace{.1in}
\theta_{1}, \theta_{2} \in [0, 2 \pi].
\end{equation}
We developed 
this set of transformations at a rather late stage of the research 
reported here, and presently have no indications that are of any
special aid in regard to the particular difficulties/challenges posed by
the HS {\it separability}-probability question. So, the 
qubit-qubit results reported below (sec.~\ref{qubit-qubit}) 
do rely essentially upon the conditions ((\ref{limits})-(\ref{limits2})) and 
the original parameterization in terms of the $z_{ij}$'s of Bloore.

Another {\it very} interesting simplifying 
parameterization---expressed in terms of 
{\it correlations} and {\it partial correlations}---can be found in the statistical/mathematical literature \cite{kurowicka,kurowicka2,kurowicka3,
joe,makhoul}. 
(In fact, the Bloore parameterization can be readily seen---in 
retrospect---to be simply a way
of decomposing a density matrix into a {\it correlation matrix} (cf. 
\cite{deVincente}), plus its diagonal entries.)
But this too seems to have
no particular enhanced value in analyzing partial transposes. 
(The cited literature also appears to be highly 
relevant to the problem of the
{\it random} generation of density matrices.)
\subsection{Previous analysis and beta function ans{\"a}tze}
In \cite{slaterPRA2}, we studied the four nonnegativity 
conditions (as well as their counterparts ---having 
completely parallel
cancellation and univariate function properties---in the 15-dimensional
case of $4 \times 4$ density matrices with, in general, {\it complex} 
entries)  
using {\it 
numerical} (primarily quasi-Monte Carlo integration) methods. We found
a  close fit to the function \cite[Figs. 3, 4]{slaterPRA2},
\begin{equation} \label{Freal}
\mathcal{S}^{approx}_{real}(\nu)=
\left(4+\frac{1}{5 \sqrt{2}}\right)
   B\left(\frac{1}{2},\sqrt{3}\right)^8
B_{\nu }\left(\frac{1}{2},\sqrt{3}\right),
\end{equation}
entering into our formula (cf. (\ref{Vsmall}), (\ref{Vbig})),
\begin{equation} \label{Vreal}
V^{HS}_{sep/real} = 2 \int_{0}^{1} \mathcal{J}_{real}(\nu) \mathcal{S}_{real}(\nu)
d \nu = \int_{0}^{\infty} \mathcal{J}_{real}(\nu) \mathcal{S}_{real}(\nu) d \nu,
\end{equation}
 for the 9-dimensional 
Hilbert-Schmidt {\it separable} volume
of the real $4 \times 4$ density matrices \cite[eq. (9)]{slaterPRA2}.
Here, $B$ denotes the (complete)
beta function, and $B_{\nu}$ the {\it incomplete} beta
function \cite{handbook},
\begin{equation} \label{betaformula}
B_{\nu}(a,b) =\int_{0}^{\nu} w^{a-1} (1-w)^{b-1} d w.
\end{equation}
Additionally \cite[eq. (10)]{slaterPRA2},
\begin{equation} \label{Jacreal}
\mathcal{J}_{real}(\nu) = \frac{\nu ^{3/2} \left(12 \left(\nu  (\nu +2) \left(\nu
   ^2+14 \nu +8\right)+1\right) \log \left(\sqrt{\nu
   }\right)-5 \left(5 \nu ^4+32 \nu ^3-32 \nu
   -5\right)\right)}{3780 (\nu -1)^9}
\end{equation}
is the 
({\it apparently} 
highly oscillatory near $\nu =1$ \cite[Fig.~1]{slaterPRA2}) 
jacobian function resulting from the transformation to the 
$\nu$ variable of the Bloore 
jacobian $(\Pi_{i=1}^{4} \rho_{ii})^{\frac{3}{2}}$.
(A referee did indicate that the apparent oscillations vanished, when he 
employed a 
Maple program using 50 digits of precision. [Only in the latest Version 6 
of Mathematica is a comparable plot feasible.] Also, 
perhaps, we should refer to $\mathcal{J}_{real}(\nu)$ as a 
{\it marginal} jacobian, since it is the result of the integration 
of a {\it three}-dimensional jacobian function over two, say
$\rho_{11}$ and $\rho_{22}$, variables.)

In the 15-dimensional {\it complex} two-qubit case, we found that the 
function
\begin{equation} \label{Fcomplex}
S^{approx}_{complex}(\nu) = \left( \frac{100000000}{2
   \sqrt[3]{2}+\frac{10^{3/4}}{3^{2/3}}} \right)
B\left(\frac{2
   \sqrt{6}}{5},\frac{3}{\sqrt{2}}\right)^{14}
B_{\nu }\left(\frac{2
   \sqrt{6}}{5},\frac{3}{\sqrt{2}}\right),
\end{equation}
provided a close fit to our numerical results \cite[eq. (14)]{slaterPRA}.
\subsection{Research design and objectives} 
Although we were able to implement the three (six-variable) 
nonnegativity conditions 
((\ref{firstcondition}), (\ref{secondcondition}) and (\ref{2by2}))
exactly in Mathematica in \cite{slaterPRA2}, 
for density matrices of the form (\ref{BlooreDenMat}),
we found that additionally incorporating the fourth
Peres-Horodecki (separability) one 
(\ref{Thirdcontion})---even
holding $\nu$ fixed at specific values---seemed to yield a 
computationally intractable problem.

In light of the apparent computational intractability in
obtaining {\it exact} results in the
9-dimensional real 
(and {\it a fortiori} 15-dimensional complex) 
two-qubit cases, we adjusted the research 
program pursued in \cite{slaterPRA2}. We now sought to 
determine how
far we would have to curtail the dimension 
(the number of free parameters) of the two-qubit systems 
 in order to be 
able to obtain {\it 
exact} results using the same basic investigative framework.
Such results---in addition to their own intrinsic interest---might 
help us understand 
those previously obtained (basically 
numerically) 
in the {\it full} 9-dimensional real 
and 15-dimensional complex cases \cite{slaterPRA2} (which, retrospectively, 
in fact, we assert does turn out to be the case).

To pursue this lower-dimensional exact strategem, 
we nullified various $m$-subsets of 
the six symmetrically-located 
off-diagonal pairs in the 9-parameter real 
density matrix (\ref{BlooreDenMat}), 
and tried to  exactly implement the so-reduced 
non-negativity conditions
((\ref{firstcondition}), (\ref{secondcondition}), (\ref{2by2}) and 
(\ref{Thirdcontion}))---both the first three (to obtain HS {\it total} 
volumes) and then all four 
jointly (to obtain HS {\it separable} volumes).
We leave the four {\it diagonal} entries themselves alone in all our 
analyses, so if we nullify
$m$ pairs of symmetically-located off-diagonal 
entries, we are left in a
(9-m)-dimensional setting.
We consider the various 
combinatorially distinct scenarios individually, though it would appear 
that we also could have grouped them into classes of scenarios equivalent under
{\it local} operations, and simply analyzed a single representative 
member of each equivalence class.

We will be examining a number of scenarios of various
 dimensionalities (that is, differing numbers of variables 
parameterizing $\rho$). 
In all of them,
we will seek to find the {\it univariate}
function $\mathcal{S}_{scenario}(\nu)$ 
(our primary computational and theoretical 
challenge) and the constant $c_{scenario}$, 
such that
\begin{equation} \label{Vsmall}
V^{HS}_{sep/scenario}= \int_{0}^{\infty} \mathcal{S}_{scenario}(\nu) \mathcal{J}_{scenario}(\nu) 
d \nu,
\end{equation}
and
\begin{equation} \label{Vbig}
V^{HS}_{tot/scenario}= c_{scenario} \int_{0}^{\infty}  \mathcal{J}_{scenario}(\nu) d \nu.
\end{equation}
Given such a pair of volumes, one can immediately
calculate the corresponding HS separability {\it probability},
\begin{equation}
P^{HS}_{sep/scenario}=\frac{V^{HS}_{sep/scenario}}{V^{HS}_{tot/scenario}}.
\end{equation}

Let us note that in the full 9-dimensional real and 15-dimensional
complex two-qubit cases 
recently studied in \cite{slaterPRA2}, it was quite natural
to expect 
that $\mathcal{S}_{real}(\nu)= \mathcal{S}_{real}(\frac{1}{\nu})$ (and
$\mathcal{S}_{complex}(\nu)= \mathcal{S}_{complex}(\frac{1}{\nu})$).
But, here, in our {\it lower}-dimensional scenarios, the nullification
of entries that we employ, breaks symmetry (duality), so we can not realistically
expect such a reciprocity property to hold, in general. 
Consequently, we adopt the more general, broader formula in  
(\ref{Vreal}) as our working formula (\ref{Vsmall}).

We now embark upon a series of multifarious lower-dimensional analyses, first 
for qubit-qubit and then qubit-qutrit, qutrit-qutrit and 
qubit-qubit-qubit systems. These will prove useful---as was our 
original hope---in developing approaches
to higher-dimensional analyses, presently out of the reach of exact
computer analyses.
\section{Qubit-Qubit Analyses} \label{qubit-qubit}
To begin, let us make the simple observation that since
the partial transposition operation on a $4 \times 4$ density matrix
interchanges only the (1,4) and (2,3) entries (and the (4,1) and (3,2) 
entries), any scenario which does not involve at least one of these
entries must only yield separable states.
\subsection{Five nullified pairs of off-diagonal entries---6 scenarios}
\subsubsection{4-dimensional real case---$P^{HS}_{sep} = 
\frac{3 \pi}{16}$} \label{subsec4dim}
There are, of course, six ways of nullifying {\it five} 
of the six off-diagonal pairs of entries
of $\rho$. Of these, only two of the six yield  any non-separable
(entangled) states. In the four trivial 
(fully separable) scenarios, the lower-dimensional counterpart to
$\mathcal{S}_{real}(\nu)$ was of the form $\mathcal{S}_{scenario}(\nu)= 
c_{scenario}=2$.

In one of the two non-trivial scenarios, having the (2,3) and (3,2) 
pair of entries 
of $\rho$ left intact (not nullified), 
the separability function was
\begin{equation} \label{equationA}
\mathcal{S}_{[(2,3)]}(\nu) =
\begin{cases}
 2 \sqrt{\nu } & 0\leq \nu \leq 1 \\
 2 & \nu >1.
\end{cases}
\end{equation}
(It is of interest to note that 
$B_{\nu}(\frac{1}{2},1) = 
2 \sqrt{\nu}$, while in \cite{slaterPRA2}, we had conjectured
that $\mathcal{S}_{real}(\nu)$ was proporitional to 
$B_{\nu}(\frac{1}{2},\sqrt{3})$.)

In the other non-trivial scenario, 
with the (1,4) and (4,1) pair being the one not nullified, the 
separability function was---in a dual manner (mapping $f(\nu)$ 
for $\nu \in [0,1]$ into $f(\frac{1}{\nu})$ for $\nu 
\in [1,\infty]$)---equal to 
\begin{equation} \label{equationB}
\mathcal{S}_{[(1,4)]}(\nu) =
\begin{cases}
 2 & 0\leq \nu \leq 1 \\
 \frac{2}{\sqrt{\nu }} & \nu >1.
\end{cases}
\end{equation}
In both of these scenarios (having $c_{scenario} =2$)
for the total (separable and non-separable) HS 
volume, we obtained 
$V^{HS}_{tot} = \frac{\pi}{48} \approx 0.0654498$
and $V^{HS}_{sep} = \frac{\pi^2}{256} \approx 0.0385531$.
The corresponding
HS separability probability for the two non-trivial (dual) scenarios 
is, then, $\frac{3 \pi}{16} \approx 0.589049$.
\subsubsection{5-dimensional complex case---$P^{HS}_{sep} 
= \frac{1}{3}$} 
\label{complex1}
Now, we allow the single
non-nullified pair of symmetrically-located entries to be {\it 
complex} in nature (so, obviously we have five variables/parameters---that 
is, including the three diagonal variables---{\it in toto} 
to consider, rather than four).

Again, we have only the same two scenarios (of the six 
combinatorially possible) being
separably non-trivial. Based on the (2,3) and (3,2) pair of entries,
the relevant function (with the slight change of notation to indicate 
complex entries) was
\begin{equation}
\mathcal{S}_{[\tilde{(2,3)}]}(\nu)= 
\begin{cases}
 \pi  \nu  & 0\leq \nu \leq 1 \\
 \pi  & \nu >1
\end{cases}
\end{equation}
and, dually, 
\begin{equation}
\mathcal{S}_{[\tilde{(1,4]})]}(\nu)= 
\begin{cases}
 \pi  & 0\leq \nu \leq 1 \\
 \frac{\pi }{\nu } & \nu >1.
\end{cases}
\end{equation}
So, the function $\sqrt{\nu}$, which appeared 
((\ref{equationA}), (\ref{equationB})) in the 
corresponding scenarios 
restricted to real entries, is replaced by $\nu$ itself in the
complex counterpart. (We note that $B_{\nu}(1,1) = \nu$.)

For both of these complex 
scenarios, we had $V^{HS}_{tot}= \frac{\pi}{120}$ and
$V^{HS}_{sep}= \frac{\pi}{360}$, for a 
particularly simple HS separability probability
of $\frac{1}{3}$.
\subsubsection{7-dimensional quaternionic  case---$P^{HS}_{sep}
= \frac{1}{10}$} \label{Dyson}
Here we allow the single pair of non-null off-diagonal entries to be
{\it quaternionic} in nature \cite{asher2,adler} \cite[sec. IV]{batle}.
We found
\begin{equation}
\mathcal{S}_{[\widetilde{(2,3)}]}(\nu)=
\begin{cases}
 \frac{\pi^2  \nu^2}{2}  & 0\leq \nu \leq 1 \\
 \frac{\pi^2}{2}  & \nu >1
\end{cases}
\end{equation}
and, dually,
\begin{equation}
\mathcal{S}_{[\widetilde{(1,4]})]}(\nu)=
\begin{cases}
 \frac{\pi^2}{2}   & 0\leq \nu \leq 1 \\
 \frac{\pi^2 }{2 \nu^2 } & \nu >1.
\end{cases}
\end{equation}
(We note that $B_{\nu}(2,1) = \frac{\nu^2}{2}$.)
For both scenarios, we had $V^{HS}_{tot}=\frac{\pi^2}{2520}, V^{HS}_{sep}=\frac{\pi^2}{25200}$, giving us
$P^{HS}_{sep}= \frac{1}{10}$---which is the {\it smallest} separability
probability we will report in this entire paper.

So, in our first set of simple ($m=5$) scenarios, we observe a decrease
in the probabilities of separability from the real to the complex to the 
quaternionic case, as well as a progression 
from $\sqrt{\nu}$ to $\nu$ to
$\nu^2$ in the functional forms occurring in the corresponding HS
separability probability functions.
\subsubsection{Relevance of Dyson indices} \label{DysonIndex}
The exponents of $\nu$ in the real-complex-quaternionic 
progression in the immediately preceding $m=5$ analyses, 
that is $\frac{1}{2},1,2$
bear an evident elementary 
relation to the {\it Dyson indices} \cite{dyson}, $\beta=1, 2, 4$,
corresponding to the Gaussian orthogonal, unitary and symplectic ensembles 
\cite{desrosiers}. (Further, many of the 
additional scenarios studied 
below---also in the non-qubit-qubit analyses---will 
have explicit occurrences in the corresponding separability functions
of $\sqrt{\nu}$ for real entries and $\nu$ for complex entries. 
Of course, use of $\mu= \sqrt{\nu}$ as our principal variable would give
the Dyson series itself, rather than one-half of it.) 
We note that 
the foundational work of \.Zyczkowski and Sommers
\cite{szHS}) in computing the HS (separable {\it plus} nonseparable) volumes 
itself 
relies strongly on random matrix theory (in particular, the Laguerre 
ensemble). Their formula for a certain generalized (Hall) normalization 
constant \cite[eq. (4.1)]{szHS}, for instance,  
contains a dummy variable $\beta$ which
equals 1 in the real case and 2 in the complex case.
In their concluding remarks, they write: ``these explicit results may be 
applied for estimation of the volume 
of the set of {\it entangled} [emphasis added] states...It is also likely
that some of the integrals obtained in this work will be useful in such
investigations'' 
\cite[p. 10125]{szHS}.

Of course, random matrix theory is framed
in terms of the eigenvalues and eigenvectors of random matrices---which
do not appear explicitly in the Bloore parameterization---so, it is 
not altogether transparent in what manner one might proceed further to 
relate the two areas.
(But for the $m=5$ highly sparse density matrices for this set of scenarios,
one can explicitly transform between the eigenvalues and the Bloore parameters.)

\subsection{Four nullified pairs of off-diagonal entries---15 scenarios}
\subsubsection{5-dimensional real case---$P^{HS}_{sep} = 
\frac{5}{8}; \frac{16}{3 \pi^2}$} \label{secd=5}
Here, there are fifteen possible scenarios, all with $V^{HS}_{tot} = 
\frac{\pi^2}{480}$. Six of them are trivial
(separability probabilities of 1), in which $c_{scenario}$
is either $\pi$ (scenarios [(1,2), (1,3)], [(1,2), (2,4)], [(1,3), (3,4)] and 
[(2,4), (3,4)]) or 4 (scenarios [(1,2), (3,4)] and [(1,3), (2,4)]).
Eight of the nine non-trivial scenarios 
all have---similarly to the 4-dimensional analyses 
(sec.~\ref{subsec4dim}) ---- separability functions $\mathcal{S}(\nu)$ 
either of the form, 
\begin{equation}
\mathcal{S}_{scenario}(\nu) = \begin{cases}
 \pi  \sqrt{\nu } & 0\leq \nu \leq 1 \\
 \pi  & \nu >1,
\end{cases}
\end{equation}
(for scenarios [(1,2), (2,3)], [(1,3), (2,3)], [(2,3), (2,4)] and
[(2,3), (3,4)])
 or, dually, 
\begin{equation}
\mathcal{S}_{scenario}(\nu) = \begin{cases}
 \pi  & 0\leq \nu \leq 1 \\
 \frac{\pi }{\sqrt{\nu }} & \nu >1
\end{cases}
\end{equation}
(for scenarios [(1,2), (1,4)], [(1,3), (1,4)], [(1,4), (2,4)] and
[(1,4), (3,4)]).
The corresponding HS
separability probabilities, for {\it 
all} eight of these non-trivial scenarios, 
are equal to 
$\frac{5}{8} = 0.625$. This result was, in all the eight cases, 
computed by taking the
the ratio of $V^{HS}_{sep} = \frac{\pi^2}{768}$ to $V^{HS}_{tot} = \frac{\pi^2}{480}$.

In the remaining (ninth) non-trivially entangled case---based on the
non-nullified dyad 
[(1,4),(2,3)]---we have, taking the ratio of $V^{HS}_{sep}= \frac{1}{90}$ to
$V^{HS}_{tot} = \frac{\pi^2}{480}$, a 
quite different Hilbert-Schmidt separability probability of 
$\frac{16}{3 \pi^2} \approx 0.54038$.
This isolated scenario (with $c_{scenario}=4$) 
can also be distinguished from the other eight 
partially entangled scenarios, in that it is the only one for
which entanglement occurs for {\it both} $\nu<1$ and $\nu>1$.
We have
\begin{equation} \label{suggestion}
\mathcal{S}_{[(1,4),(2,3)]}(\nu) = \begin{cases}
 4 \sqrt{\nu } & 0\leq \nu \leq 1 \\
 \frac{4}{\sqrt{\nu }} & \nu >1.
\end{cases}
\end{equation}
By way of illustration, in this specific case, we  have 
the scenario-specific marginal jacobian function,
\begin{equation}
\mathcal{J}_{[(1,4),(2,3)]}(\nu) = 
-\frac{\sqrt{\nu } \left(-3 \nu ^2+(\nu  (\nu +4)+1) \log
   (\nu )+3\right)}{30 (\nu -1)^5}.
\end{equation}
\subsubsection{6-dimensional {\it mixed} (real and complex) case ---$P^{HS}_{sep} = \frac{105 \pi}{512}; \frac{135 \pi}{1024}; \frac{3}{8}$}
Here, we again nullify all but two of the off-diagonal entries ($m=4$) of 
$\rho$, 
but allow the {\it first} of the two non-nullified entries to be
{\it complex} in nature. 
Making (apparently necessary) use of the
circular/trigonometric  transformation
$\rho_{11} = r^2 \sin{\theta}^2, \rho_{22} = r^2 \cos{\theta}^2$, we were
able to obtain an interesting variety of exact results.
One of these takes the form,
\begin{equation}
\mathcal{S}_{[\tilde{(1,2)},(1,4)]}(\nu)=
\mathcal{S}_{[\tilde{(1,3)},(1,4)]}(\nu)=
\begin{cases}
 \left\{\frac{4 \pi }{3},0\leq \nu \leq 1\right\} &
   \left\{\frac{4 \pi }{3 \sqrt{\nu }},\nu >1\right\}.
\end{cases}
\end{equation}
Now, we have $V^{HS}_{tot}= \frac{\pi^2}{1440}$ and
$V^{HS}_{sep} = \frac{7 \pi^3}{49152}$, so
$P^{HS}_{sep}= \frac{105 \pi}{512} \approx 0.644272$.
The two dual scenarios---having the same three results---are 
$[\tilde{(1,2)},(2,3)]$ and 
$[\tilde{(1,3)},(2,3)]$.

Additionally, we have an isolated scenario,
\begin{equation} \label{firstisolated}
\mathcal{S}_{[\tilde{(1,4)},(2,3)]}(\nu) =  \begin{cases}
 \left\{2 \pi  \sqrt{\nu },0\leq \nu \leq 1\right\} &
   \left\{\frac{2 \pi }{\nu },\nu >1\right\},
\end{cases}
\end{equation}
for which, $V^{HS}_{tot} = \frac{\pi^2}{1440}$ and
$V^{HS}_{sep}= \frac{3 \pi^3}{32768}$, so
$P^{HS}_{sep}= \frac{135 \pi}{1024} \approx 0.414175$.
(Note the presence of {\it both} $\sqrt{\nu}$ and $\nu$ in (\ref{firstisolated})---apparently related to the mixed [real and complex] nature of this 
scenario (cf. (\ref{secondmixed})).)

Further,
\begin{equation}
\mathcal{S}_{[\tilde{(1,4)},(2,4)]}(\nu)=
\mathcal{S}_{[\tilde{(1,4)},(3,4)]}(\nu)= \begin{cases}
 \left\{\frac{4 \pi }{3},0\leq \nu \leq 1\right\} &
   \left\{\frac{4 \pi }{3 \nu },\nu >1\right\},
\end{cases}
\end{equation}
the dual scenarios being $[\tilde{(2,3)},(2,4)]$ and $[\tilde{(2,3)},(3,4)]$.
For all four of these scenarios, $V^{HS}_{tot}= \frac{\pi^2}{1440}$ and
$V^{HS}_{sep}= \frac{\pi^2}{3840}$,
so $P^{HS}_{sep} =\frac{3}{8} =0.375$.
\subsubsection{7-dimensional complex
case---$P^{HS}_{sep} = \frac{2}{5}$}
Here, in an $m=4$ setting, 
we nullify four of the six off-diagonal pairs of the $4 \times 4$ 
density matrix, allowing the remaining two pairs {\it both} to be complex.
We have (again observing a shift from $\sqrt{\nu}$ in the real case
to $\nu$ in the complex case)
\begin{equation} \label{complex7}
\mathcal{S}_{[\tilde{(1,2)},\tilde{(1,4)}]}(\nu) =\mathcal{S}_{[\tilde{(1,3)},\tilde{(1,4)}]}(\nu)= 
\mathcal{S}_{[\tilde{(1,4)},\tilde{(2,4)}]}(\nu)=
\mathcal{S}_{[\tilde{(1,4)},\tilde{(3,4)}]}(\nu) =
\begin{cases}
 \frac{\pi ^2}{2} & 0\leq \nu \leq 1 \\
 \frac{\pi ^2}{2 \nu } & \nu >1.
\end{cases}
\end{equation}
Since $V^{HS}_{tot}= \frac{\pi^2}{5040}$ and
$V^{HS}_{sep}= \frac{\pi^2}{12600}$, we have
$P^{HS}_{sep}= \frac{2}{5} = 0.4$.
We have the same three outcomes for the four dual 
scenarios $[\tilde{(1,2)},\tilde{(2,3)}],
[\tilde{(1,3)},\tilde{(2,3)}],
[\tilde{(2,3)},\tilde{(2,4)}]$ and 
$[\tilde{(2,3)},\tilde{(3,4)}]$, as well as---rather remarkably---for 
the (again isolated 
[cf. (\ref{firstisolated})])
scenario $[\tilde{(1,4)},\tilde{(2,3)}]$, having the (somewhat different)
separability function (manifesting entanglement for both
$\nu<1$ and $\nu > 1$),
\begin{equation} \label{secondmixed}
\mathcal{S}_{[\tilde{(1,4)},\tilde{(2,3)}]}(\nu)  = 
\begin{cases}
 \pi ^2 \nu & 0\leq \nu \leq 1 \\
 \frac{\pi ^2}{\nu } & \nu >1.
\end{cases}
\end{equation}
(However, $c_{scenario} = \pi^2$ for this isolated scenario, while it
equals
$\frac{\pi^2}{2}$ for the other eight.)
The remaining six (fully separable) scenarios (of the fifteen possible)
simply have $P^{HS}_{sep}=1$.
\subsubsection{8-dimensional mixed (real and quaternionic) case}
We report here that
\begin{equation}
c_{[\widetilde{(1,2)},(1,4)]} =\frac{8 \pi^2}{15}, \hspace{.2in} 
c_{[(1,2),\widetilde{(1,4)}]}= 32,
\end{equation}
where as before the wide tilde notation denotes the quaternionic 
off-diagonal entry.
\subsection{Three nullified pairs of off-diagonal entries---20 scenarios}
\subsubsection{6-dimensional 
real case---$P^{HS}_{sep} =2 -\frac{435 \pi}{1024}$}
Here ($m=3$), 
there are twenty possible scenarios---nullifying {\it triads} of
off-diagonal pairs in $\rho$.
Of these twenty, there are 
four totally separable scenarios---corresponding to the non-nullified triads [(1,2), (1,3), (2,4)], [(1,2), (1,3), (3,4)], [(1,2), (2,4), (3,4)] and [(1,3), (2,4), (3,4)]---with $c_{scenario} = \frac{\pi^2}{2}$ and 
$V^{HS}_{tot} = V^{HS}_{sep} = \frac{\pi^3}{5760}$.
To proceed further in this 6-dimensional case---in which we 
began to 
encounter some computational difficulties---we sought, again,  to 
enforce the four  nonnegativity conditions
((\ref{firstcondition}), (\ref{secondcondition}), (\ref{2by2}),
(\ref{Thirdcontion})), but only after setting $\nu$ to specific values,
rather than allowing $\nu$ to vary.
We chose the nine values
$\nu= \frac{1}{5}, \frac{2}{5}, \frac{3}{5}, \frac{4}{5}$, 1, 2, 3, 4 and 5.
Two of the scenarios (with the triads [(1,2), (2,3), (3,4)] and
[(1,3),(2,3),(2,4)]) could, then, be seen to fit unequivocally  
into our earlier observed 
predominant pattern,
having the piecewise separability function,
\begin{equation} \label{1case}
\mathcal{S}_{[(1,2), (2,3), (3,4)]}(\nu) = 
\mathcal{S}_{[(1,3), (2,3), (2,4)]}(\nu) =
\begin{cases}
 \frac{\pi ^2 \sqrt{\nu }}{2} & 0\leq \nu \leq 1 \\
 \frac{\pi ^2}{2 } & \nu >1.
\end{cases}
\end{equation}
We, then, computed for these two scenarios that 
$V^{HS}_{tot} = \frac{\pi^3}{5760} \approx 0.00538303$ and 
(again making use of the transformation
$\rho_{11} = r^2 \sin{\theta}^2, \rho_{22} = r^2 \cos{\theta}^2$) that 
$V^{HS}_{sep} = 2 \left(\frac{\pi ^3}{5760}-\frac{29 \pi^4}{786432}\right) 
\approx 0.00358207$. This gives  us 
$P^{HS}_{sep} = 2-\frac{435 \pi }{1024} \approx 0.665437$.
For two dual dyads, we have the same volumes and separability 
probability and, now, the piecewise separability function,
\begin{equation} \label{onecase}
\mathcal{S}_{[(1,2), (1,4), (3,4)]}(\nu)= \mathcal{S}_{[(1,3), (1,4), (2,4)]}(\nu) =
\begin{cases}
 \frac{\pi ^2}{2 } & 0\leq \nu \leq 1 \\
 \frac{\pi ^2}{2 \sqrt{\nu}} & \nu >1.
\end{cases}
\end{equation}

We have not, to this point, been able to explicitly 
and succinctly characterize
the functions $\mathcal{S}_{scenario}(\nu)$ for non-trivial 
fully real $m=3$ scenarios 
other than the dual pair 
((\ref{1case}), (\ref{onecase})).

In all the separably non-trivial scenarios so far presented and discussed,
we have had the relationship $\mathcal{S}_{scenario}(1) = c_{scenario}$. However,
in our present $m=3$ setting (three pairs of 
nullified off-diagonal entries), we have
situations in which $\mathcal{S}_{scenario}(1) <  c_{scenario}$.
The values of $c_{scenario}$ in the sixteen non-trivial 
fully real $m=3$ scenarios 
are either $\frac{\pi^2}{2} \approx 4.9348$ (twelve occurrences) or
$\frac{4 \pi}{3} \approx 4.18879$ (four occurrences---[(1,2), (1,3), (1,4)],  [(1,2), (2,3), (2,4)], [(1,3), (2,3), (3,4)] and [(1,4), (2,4), (3,4)]). 
In all four of 
the latter ($\frac{4 \pi}{3}$) occurrences, though, 
we have the {\it inequality},
\begin{equation}
\mathcal{S}_{scenario}(1) = 
\frac{1}{24} \left(12+16 \pi +3 \pi ^2\right)
\approx 3.8281 < \frac{4 \pi}{3} \approx 4.18879,
\end{equation}
as well as a parallel 
inequality for four of the twelve former ($\frac{\pi^2}{2}$) 
cases.
The implication of these inequalities for those eight 
scenarios is that at $\nu=1$ (the value associated with the fully
mixed [separable] classical state), 
that is, when $\rho_{11} \rho_{44} = \rho_{22} \rho_{33}$, there do exist
non-separable states.

\subsubsection{7-dimensional mixed (one complex and two real) case---
$P_{sep}^{HS} = \frac{11}{16}$}
Here, in an $m=3$ setting, we take the first entry of the 
non-nullified triad to be 
complex and the other two real. 
Of the twenty possible scenarios, four ---- $[\tilde{(1,2)},(1,3),(1,4)], 
[\tilde{(1,2)},(2,3),(2,4)],
[\tilde{(1,3)},(2,3),(3,4)]$ and 
$[\tilde{(1,4)},(2,4),(3,4)]$---had $c_{scenario}=\frac{\pi^2}{2} 
\approx 4.9348$ and these four all had the same (lesser) value of
\begin{equation} \label{Sscenario}
\mathcal{S}_{scenario}(1) = \frac{56}{27}+\frac{\pi^2}{4} \approx 4.54148.
\end{equation}

There were seven scenarios with $c_{scenario} = \frac{16 \pi}{9} 
\approx 5.58505$. Three of them---$[\tilde{(1,2)},(1,3),(2,4)],
[\tilde{(1,2)},(1,4),(2,3)]$ and 
$[\tilde{(1,3)},(1,4),(2,3)]$ ---had
$\mathcal{S}_{scenario}(1)=  
\frac{16 \pi}{9}$ (manifesting equality), 
while four---$[\tilde{(1,2)},(1,3),(2,3)],
[\tilde{(1,2)},(1,4),(2,4)],
[\tilde{(1,3)},(1,4),(3,4)]$ and
$[\tilde{(2,3)},(2,4),(3,4)]$---had the result 
(\ref{Sscenario}) (manifesting inequality).

The remaining nine of the twenty scenarios all had $c_{scenario} = 
\mathcal{S}_{scenario}(1) = \frac{2 \pi^2}{3} \approx 6.57974$.
For one of them, we obtained 
\begin{equation}
\mathcal{S}_{[\tilde{(1,2)},(2,3),(3,4)]}(\nu) = \begin{cases}
 \frac{2 \pi ^2}{3} & \nu \geq 1 \\
 \frac{2 \pi ^2 \sqrt{\nu }}{3} & 0<\nu <1,
\end{cases}
\end{equation}
with associated values of $V_{tot}^{HS}= \frac{\pi^3}{20160}$,
$V_{sep}^{HS}= \frac{11 \pi^3}{322560}$ and
$P_{sep}^{HS}= \frac{11}{16} \approx 0.6875$.
A dual scenario to this one that we were able to find was 
$[\tilde{(1,2)},(1,4),(3,4)]$.
The separability functions---and, hence, separability 
probabilities---for the other eighteen scenarios, however, are unknown
to us at present.
\subsubsection{8-dimensional mixed (two complex and one real)  case}
Our sole result in this category is
\begin{equation}
c_{[\tilde{(1,2)},\tilde{(1,3)},(1,4)]}=\frac{8 \pi^2}{15}.
\end{equation}
\subsubsection{9-dimensional complex case}
Now, we have three off-diagonal complex entries, requiring {\it six}
parameters for their specification. 
This is about the limit in the number
of free off-diagonal parameters for which we might
hopefully be able to determine associated separability functions.

As initial findings, we obtained
\begin{equation}
\mathcal{S}_{[\tilde{(1,4)},\tilde{(2,3)},\tilde{(2,4)}]}(1) = 
c_{[\tilde{(1,4)},\tilde{(2,3)},\tilde{(2,4)}]} =
\frac{\pi^3}{4},
\end{equation}
and also for scenarios $[\tilde{(1,2)},\tilde{(1,4)},\tilde{(3,4)}], 
[\tilde{(1,3)},\tilde{(1,4)},\tilde{(2,4)}]$ and
$[\tilde{(1,4)},\tilde{(2,3)},\tilde{(3,4)}]$,
while
\begin{equation}
c_{[\tilde{(1,2)},\tilde{(1,3)},\tilde{(1,4)}]} = 
c_{[\tilde{(1,3)},\tilde{(2,3)},\tilde{(3,4)}]} =
c_{[\tilde{(1,4)},\tilde{(2,4)},\tilde{(3,4)}]} =
\frac{\pi^3}{6}.
\end{equation}
\subsection{Two or fewer nullified pairs of off-diagonal entries}
\subsubsection{7-dimensional real case}
The [(1,2), (1,3), (2,4), (3,4)] scenario is
the only fully separable one
of the fifteen possible ($m=2$). For all 
the other fourteen non-trivial scenarios,
there are non-separable states {\it both} for $\nu<1$ and $\nu>1$.
For all fifteen scenarios, we have $c_{scenario} = \frac{2 \pi^2}{3} 
\approx 6.57974$.
Otherwise, we 
have not so far been able to extend the analyses above to this $m=2$ 
fully real
case (and {\it a fortiori} the $m=1$ fully real case), 
even to determine specific values of $\mathcal{S}_{scenario}(1)$.
\subsubsection{8-dimensional real case} \label{sec8dim}
Here we have $c_{scenario} = \frac{8 \pi^2}{9} \approx 
8.77298$ for all the six possible (separably non-trivial)
scenarios ($m=1$). Let us note that this is, in terms of preceding values of 
these constants (for the successively lower-dimensional fully real 
scenarios), $\frac{8 \pi^2}{9} = \frac{4}{3} (\frac{2 \pi^2}{3})$, while
$\frac{2 \pi^2}{3} = \frac{4}{3} (\frac{\pi^2}{2})$. 
Also, 
$\frac{32 \pi^2}{27} = \frac{4}{3} (\frac{8 \pi^2}{9}$), the 
further relevance of
which will be apparent in relation to our discussion of 
the full 9-dimensional real scenario (sec.~\ref{twoqubitconjectures}).

\section{Qubit-Qutrit Analyses} \label{qubitqutrit}
The cancellation property, we exploited above,  of the Bloore
parameterization---by which the determinant and 
principal minors of density matrices
can be {\it factored} 
into products of (nonnegative) diagonal entries and terms just 
involving off-diagonal parameters ($z_{ij}$)---clearly 
extends to $n \times n$ density matrices. It initially appeared to us that the advantage of the 
parameterization in studying the 
two-qubit HS separability probability question would diminish if one
were to examine the two-qubit separability problem for other (possibly 
{\it monotone}) metrics than the HS one (cf. \cite{BuresSep}), 
or even the qubit-{\it qutrit} {\it HS} separability
probability question. 
But upon some further analysis, we have found that the 
nonnegativity condition for the
determinant of the partial transpose of a real $6 \times 6$ (qubit-qutrit)
density matrix (cf. (\ref{firstcondition}))
can be expressed in terms of the corresponding $z_{ij}$'s
and {\it two} ratio variables (thus, not requiring the five independent
diagonal variables individually), 
\begin{equation} \label{tworatios}
\nu_{1}= \frac{\rho_{11} \rho_{55}}{\rho_{22} \rho_{44}}, \hspace{.2in}
\nu_{2}= \frac{\rho_{22} \rho_{66}}{\rho_{33} \rho_{55}},
\end{equation}
rather than simply one ($\nu$) as in the
$4 \times 4$ case. 
(We compute the qubit-qutrit 
partial transpose by transposing in place the four
$3 \times 3$ blocks of $\rho$, rather than---as we might 
alternatively have done---the nine $2 \times 2$ blocks.)
\subsection{Fourteen nullified pairs of off-diagonal entries---15 scenarios}
\subsubsection{6-dimensional real case---$P^{HS}_{sep} =\frac{3 \pi}{16}$}
To begin our examination of  the qubit-qutrit case, we study the 
($m=14$) scenarios, 
in which only a single pair of real entries is left 
intact and all other off-diagonal pairs of the $6 \times 6$ 
density matrix are nullified. (We not only require that the determinant
of the partial transpose of $\rho$ be nonnegative 
for separability to hold---as suffices 
in the qubit-qubit case, given that $\rho$ itself is a density 
matrix  \cite{ver,augusiak}---but also, {\it per} the 
Sylvester criterion,  a nested series of 
principal leading minors of $\rho$.)
We have six separably non-trivial scenarios. 
(For all of them, $V_{tot}^{HS}= \frac{\pi}{1440}$.)

Firstly, we
have the separability
function,
\begin{equation} \label{S15}
\mathcal{S}_{[(1,5)]}^{6 \times 6}(\nu_{1})=
\begin{cases}
 2 & \nu _1\leq 1 \\
 \frac{2}{\sqrt{\nu _1}} & \nu_{1}>1.
\end{cases}
\end{equation}
The dual scenario to this is [(2,4)].
Further,
\begin{equation}
\mathcal{S}_{[(1,6)]}^{6 \times 6}(\nu_{1},\nu_{2})=
\begin{cases}
 2 & \nu_{1} \nu _2 \leq 1 \\ 
 \frac{2}{\sqrt{\nu _1 \nu _2}} & \nu _1 \nu_2 >  1,
\end{cases}
\end{equation}
with the dual scenario here being [(3,4)].
Finally,
\begin{equation}
\mathcal{S}_{[(2,6)]}^{6 \times 6}(\nu_{2})=
\begin{cases}
 2 & \nu _2\leq 1 \\
 \frac{2}{\sqrt{\nu _2}} & \nu_{2} > 1,
\end{cases}
\end{equation}
having the dual [(3,5)].

The remaining nine possible scenarios---the same as their 
complex counterparts in the immediate
next analysis---are all fully separable in character.

We have found that $V^{HS}_{sep} = \frac{\pi^2}{7680}$ for the
six non-trivially separable scenarios here, so $P^{HS}_{sep} = 
\frac{3 \pi}{16} \approx 0.589049$, as in the qubit-qubit analogous
case (sec.~\ref{subsec4dim}).
\subsubsection{7-dimensional complex case---$P^{HS}_{sep}= \frac{1}{3}$}  \label{ComplexSection}
Now, we allow the single non-nullified pair of off-diagonal entries
to be complex in nature (the two paired entries, of course, being complex
conjugates of one another). ($V_{tot}^{HS}= \frac{\pi}{5040}$ for this
series of fifteen scenarios.)
Then, we have (its dual being $[\tilde{(2,4)}]$)
\begin{equation}
\mathcal{S}_{[\tilde{(1,5)}]}^{6 \times 6}(\nu_{1})=
\begin{cases}
 \pi  & \nu _1\leq 1 \\
 \frac{\pi }{\nu _1} & \nu_{1} > 1.
\end{cases}
\end{equation}

Further, we have  (with the dual $[\tilde{(3,4)}]$)
\begin{equation}
\mathcal{S}_{[\tilde{(1,6)}]}^{6 \times 6}(\nu_{1},\nu_{2})=
\begin{cases}
 \pi  & \nu_{1} \nu _2 \leq 1 \\ 
 \frac{\pi }{\nu _1 \nu _2} & \nu _1 \nu_{2} > 1
\end{cases}
\end{equation}
and (its dual being $[\tilde{(3,5)}]$),
\begin{equation}
\mathcal{S}_{[\tilde{(2,6)}]}^{6 \times 6}(\nu_{2})=
\begin{cases}
 \pi  & \nu _2\leq 1 \\
 \frac{\pi }{\nu _2} & \nu_{2} > 1.
\end{cases}
\end{equation}
For all six of these scenarios,
$V_{sep}= \frac{\pi}{15120}$, so $P_{HS}^{sep} = \frac{1}{3}$.
\subsection{Thirteen nullified pairs of off-diagonal entries---105 scenarios}
\subsubsection{7-dimensional real case---$P^{HS}_{sep}=\frac{5}{8}; \frac{5}{16}; \frac{3 \pi}{32}; \frac{16}{3 \pi^2}$}
Continuing along similar lines ($m=13$), we have 105 combinatorially
distinct possible scenarios. 
Among the separably non-trivial scenarios, we have
\begin{equation}
\mathcal{S}_{[(1,2),(1,5)]}^{6 \times 6}(\nu_1)=
\mathcal{S}_{[(1,4),(1,5)]}^{6 \times 6}(\nu_1)=
\begin{cases}
 \pi  & \nu _1\leq 1 \\
 \frac{\pi }{\sqrt{\nu _1}} & \nu_{1} > 1,
\end{cases}
\end{equation}
(duals being [(1,2),(2,4)] and [(1,4),(2,4)]).  We computed  $V^{HS}_{tot}= 
\frac{\pi^2}{20610}, V^{HS}_{sep}= \frac{\pi^2}{32256}$, so
$P^{HS}_{sep}=\frac{5}{8} = 0.625$ for these scenarios.

Also,
\begin{equation}
\mathcal{S}_{[(1,3),(1,6)]}^{6 \times 6}(\nu_{1},\nu_{2}) =
\mathcal{S}_{[(1,4),(1,6)]}^{6 \times 6}(\nu_{1},\nu_{2}) =
\begin{cases}
 \pi  & \nu_{1} \nu_{2} < 1  \\
 \frac{\pi }{\sqrt{\nu _1} \sqrt{\nu _2}} & \nu_{1} \nu_{2} \geq  1.
\end{cases}
\end{equation}
We, then,  have $V^{HS}_{tot}=
\frac{\pi^2}{20610}, V^{HS}_{sep}= \frac{\pi^2}{64512}$, so
$P^{HS}_{sep}=\frac{5}{16} = 0.3125$.

Additionally,
\begin{equation}
\mathcal{S}_{[(1,4),(2,6)]}^{6 \times 6}(\nu_{2}) =
\begin{cases}
 4 & \nu _2\leq 1 \\
 \frac{4}{\sqrt{\nu _2}} & \nu_{2} > 1.
\end{cases}
\end{equation}
For this scenario, we have 
$V^{HS}_{tot}=
\frac{\pi^2}{20610}, V^{HS}_{sep}= \frac{\pi^2}{215040}$, so
$P^{HS}_{sep}=\frac{3 \pi}{32} \approx 0.294524$.

Further still,
\begin{equation}
\mathcal{S}_{[(1,5),(2,4)]}^{6 \times 6}(\nu_{1}) =
\begin{cases}
 \frac{4}{\sqrt{\nu _1}} & \nu _1>1 \\
 4 \sqrt{\nu _1} & \nu_{1} \leq 1.
\end{cases}
\end{equation}
For this scenario, we have
$V^{HS}_{tot}=
\frac{\pi^2}{20610}, V^{HS}_{sep}= \frac{1}{3780}$, so
$P^{HS}_{sep}=\frac{16}{3 \pi^2} \approx 0.54038$. 

Further,
\begin{equation}
\mathcal{S}_{[(1,2),(2,6)]}^{6 \times 6}(\nu_{2})=
\begin{cases}
 \pi  & \nu _2\leq 1 \\
 2 \left(\cos ^{-1}\left(\sqrt{1-\frac{1}{\nu
   _2}}\right)+\frac{\sqrt{\nu _2-1}}{\nu _2}\right) &
   \nu_{2} > 1.
\end{cases}
\end{equation}
The separability function for [(1,3),(1,5)] is obtained from this one
by replacing $\nu_{2}$ by $\nu_{1}$.

Also,
\begin{equation}
\mathcal{S}^{6 \times 6}_{[(1,2),(3,4)]}(\nu_{1},\nu_2) =
\end{equation}
\begin{displaymath}
\begin{cases}
\pi \sqrt{\nu_{1}} \sqrt{\nu_{2}} & \nu_{1} \nu_{2} <1 \\
 4 \sqrt{1-\frac{1}{\nu _1 \nu _2}}-\frac{2 \left(i \log
   \left(\frac{\sqrt{\nu _1 \nu _2-1}+i}{\sqrt{\nu _1}
   \sqrt{\nu _2}}\right) \nu _1 \nu _2+\sqrt{\nu _1 \nu
   _2-1}\right)}{\sqrt{\nu _1} \sqrt{\nu _2}} & \nu_{1} \nu_{2} \geq 1.
\end{cases}
\end{displaymath}
The separability function for [(1,2),(3,5)] 
can be obtained from this one by setting $\nu_{1}=1$.
\subsubsection{8-dimensional mixed (real and complex) case---$P^{HS}_{sep} = \frac{105 \pi}{512}$}
Further, we have (with $V_{tot}^{HS} = 
\frac{\pi^2}{80640}$ for all scenarios),
\begin{equation} \label{firstmixed}
\mathcal{S}_{[\tilde{(1,2)},(2,4)]}^{6 \times 6}(\nu_{1})=
\mathcal{S}_{[\tilde{(1,4)},(2,4)]}^{6 \times 6}(\nu_{1})=
\begin{cases}
 \frac{4 \pi }{3} & \nu _1\geq 1  \\
 \frac{4 \pi  \sqrt{\nu _1}}{3} & 0<\nu _1<1.
\end{cases}
\end{equation}
Since $V^{HS}_{sep} = \frac{\pi^3}{393216}$, we have
$P^{HS}_{sep} = \frac{105 \pi}{512} \approx 0.644272$ for both 
these scenarios.

Further,
\begin{equation}
\mathcal{S}_{[\tilde{(1,3)},(2,4)]}^{6 \times 6}(\nu_{1})=
\begin{cases}
 \frac{4 \pi  \sqrt{\nu _1}}{3} & 0<\nu _1 \leq 1 \\
 2 \pi -\frac{2 \pi }{3 \nu _1} & \nu _1>1
\end{cases}
\end{equation}
and
\begin{equation}
\mathcal{S}_{[\tilde{(1,3)},(3,4)]}^{6 \times 6}(\nu_{1},\nu_{2})=
\mathcal{S}_{[\tilde{(1,4)},(3,4)]}^{6 \times 6}(\nu_{1},\nu_{2})=
\begin{cases}
 \frac{4 \pi }{3} & \nu _1 \nu _2\geq 1 \\
 \frac{4}{3} \pi  \sqrt{\nu _1 \nu _2} & 0<\nu _1 \nu _2<1.
\end{cases}
\end{equation}

Additionally,
\begin{equation} \label{puzzle}
\mathcal{S}_{[\tilde{(2,3)},(3,4)]}^{6 \times 6}(\nu_{1},\nu_{2})= 
\begin{cases}
 \frac{4 \pi }{3} & \nu _1 \nu _2\geq 1 \\
 \frac{2}{3} \pi  \sqrt{\nu _1 \nu _2} \left(3-\nu _1 \nu_2  \right) & 0<\nu _1 \nu _2<1
\end{cases}
\end{equation}
and
\begin{equation}
\mathcal{S}_{[\tilde{(1,2)},(3,4)]}^{6 \times 6}(\nu_{1},\nu_{2})=
\begin{cases}
 \frac{4 \pi }{3} & \nu _1 \nu _2=1 \\
 \frac{4}{3} \pi  \sqrt{\nu _1 \nu _2} & 0<\nu _1 \nu _2<1
   \\
 \pi  \left(\sqrt{\nu _1 \nu _2}+2\right)-\frac{2 \pi }{3
   \nu _1 \nu _2} & \nu _1 \nu _2>1.
\end{cases}
\end{equation}

\begin{figure}
\includegraphics{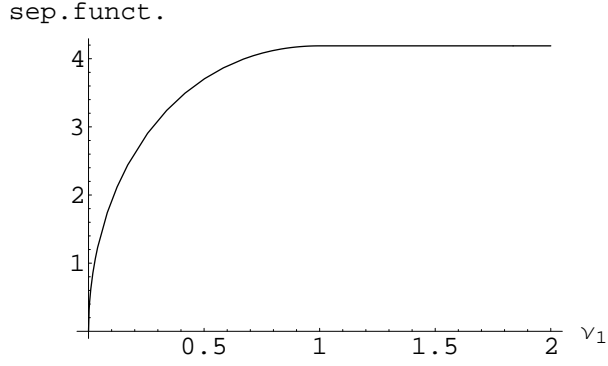}
\caption{\label{fig:S2324}Plot of the separability function
$S_{[\tilde{(2,3)},(2,4)]}^{6 \times 6}(\nu_{1})$}
\end{figure}

We have also obtained the separability function
(Fig.~\ref{fig:S2324})
\begin{equation} \label{morepuzzle}
\mathcal{S}_{[\tilde{(2,3)},(2,4)]}^{6 \times 6}(\nu_{1})=
\begin{cases}
 \frac{4 \pi }{3} & \nu _1\geq 1 \\
 \frac{2}{3} \pi  \left(3 -\nu _1 \right) \sqrt{\nu _1} &
   0<\nu _1<1.
\end{cases}
\end{equation}
Of the 105 possible scenarios, sixty had $S^{6 \times 6}_{scenario}(1,1) = 
c_{scenario} = \frac{4 \pi}{3}$, thirty-three 
had $S^{6 \times 6}_{scenario}(1,1) =
c_{scenario} = 2 \pi$, and twelve (for example, $[\tilde{(3,4)},(5,6)]$) had
$S^{6 \times 6}_{scenario}(1,1) = \frac{4 \pi}{3} < c_{scenario} = 2 \pi$.
\subsubsection{9-dimensional complex case---$P^{HS}_{sep} = \frac{1}{3}; \frac{2}{5}$}
We have obtained the results
\begin{equation}
\mathcal{S}_{[\tilde{(1,2)},\tilde{(2,4)}]}^{6 \times 6}(\nu_{1})= 
\mathcal{S}_{[\tilde{(1,4)},\tilde{(2,4)}]}^{6 \times 6}(\nu_{1})=
\begin{cases}
 \frac{\pi ^2 \nu_{1} }{2} & 0\leq \nu_{1} \leq 1 \\
 \frac{\pi ^2}{2 } & \nu_{1} >1.
\end{cases}
\end{equation}
Since $V^{HS}_{tot} = \frac{\pi^2}{362880}$ and
$V^{HS}_{sep}= \frac{\pi^2}{907200}$, we have here
$P^{HS}_{sep} = \frac{2}{5} = 0.4$. 
We have the same three outcomes also based on the separability function,
\begin{equation}
\mathcal{S}_{[\tilde{(1,4)},\tilde{(3,4)}]}^{6 \times 6}(\nu_{1},\nu_{2})=
\begin{cases}
 \frac{\pi ^2}{2} & \nu _1 \nu _2\geq 1 \\
 \frac{1}{2} \pi ^2 \nu _1 \nu _2 & \nu _1 \nu _2<1.
\end{cases}
\end{equation}

Further,
\begin{equation}
\mathcal{S}_{[\tilde{(1,2)},\tilde{(3,4)}]}^{6 \times 6}(\nu_{1},\nu_{2})=
\begin{cases}
 \frac{2 \pi ^2 \nu _1 \nu _2-\pi ^2}{2 \nu _1 \nu _2} &
   \nu _1 \nu _2>1 \\
 \frac{1}{2} \pi ^2 \nu _1 \nu _2 & 0<\nu _1 \nu _2 \leq 1.
\end{cases}
\end{equation}
and
\begin{equation}
\mathcal{S}_{[\tilde{(1,3)},\tilde{(2,4)}]}^{6 \times 6}(\nu_{1})=
\begin{cases}
 \frac{\pi ^2 \nu _1}{2} & 0<\nu _1 \leq 1 \\
 \pi ^2-\frac{\pi ^2}{2 \nu _1} & \nu _1>1.
\end{cases}
\end{equation}
For both of these last two scenarios, we have $V_{tot}^{HS}= \frac{\pi^2}{362880}$ and 
$V^{HS}_{sep}= \frac{\pi^2}{362880}$, leading to
$P^{HS}_{sep}= \frac{1}{3} \approx 0.33333$.
Also, we have these same three outcomes based on the 
separability function,
\begin{equation} \label{puzzle2}
\mathcal{S}_{[\tilde{(2,3)},\tilde{(3,4)}]}^{6 \times 6}(\nu_{1},\nu_{2})=
\begin{cases}
 \frac{\pi ^2}{2} & \nu _1 \nu _2\geq 1 \\
 \frac{1}{2} \pi ^2 \nu _1 \nu _2 \left(2 - \nu _1 \nu
   _2\right) & 0<\nu _1 \nu _2<1.
\end{cases}
\end{equation}

Of the 105 possible scenarios---in complete parallel to those 
in the immediately preceding section---sixty 
had $S^{6 \times 6}_{scenario}(1,1) =
c_{scenario} = \frac{\pi^2}{2}$, thirty-three
had $S^{6 \times 6}_{scenario}(1,1) =
c_{scenario} = \pi^2$, and twelve (for example, 
$[\tilde{(3,4)},\tilde{(5,6)}]$) had
$S^{6 \times 6}_{scenario}(1,1) = \frac{\pi^2}{2} < c_{scenario} =  \pi^2$.

Our results in this (9-dimensional) section and the 
(8-dimensional) one immediately preceding it are still incomplete 
with respect to various scenario-specific separability functions
and, thus, the associated HS separability properties.
\subsection{Twelve nullified pairs of off-diagonal entries---455 scenarios}
\subsubsection{8-dimensional real case}
Now, we allow three of the off-diagonal pairs of entries to be non-zero,
but also require them to be simply real.
We found the separability function
\begin{equation}
\mathcal{S}_{[(1,2),(1,3),(3,4)]}(\nu_{1},\nu_{2}) = 
\mathcal{S}_{[(1,2),(1,4),(3,4)]}(\nu_{1},\nu_{2}) =
\end{equation}
\begin{displaymath}
\begin{cases}
 \frac{4 \pi }{3} & \frac{1}{\nu _1}=\nu _2\land \nu _1>0
   \\
 \frac{4}{3} \pi  \sqrt{\nu _1 \nu _2} & \nu _1>0\land
   \frac{1}{\nu _1}>\nu _2\land \nu _2>0 \\
 \frac{1}{3} \pi  \left(3 \sec ^{-1}\left(\sqrt{\nu _1 \nu
   _2}\right)+4 \sqrt{\nu _1 \nu _2}+\frac{\sqrt{\nu _1
   \nu _2-1}}{\nu _1 \nu _2}-4 \sqrt{\nu _1 \nu
   _2-1}\right) & \nu _1>0\land \frac{1}{\nu _1}<\nu _2.
\end{cases}
\end{displaymath}
Also, we have
\begin{equation}
\mathcal{S}_{[(1,3),(1,4),(2,4)]}(\nu_{1}) =
\begin{cases}
 \frac{4 \pi }{3} & \nu _1=1 \\
 \frac{4 \pi  \sqrt{\nu _1}}{3} & 0<\nu _1<1 \\
 \frac{\pi  \left(4 \nu _1^{3/2}+\left(3 \sin
   ^{-1}\left(\sqrt{1-\frac{1}{\nu _1}}\right)-4 \sqrt{\nu
   _1-1}\right) \nu _1+\sqrt{\nu _1-1}\right)}{3 \nu _1} &
   \nu _1>1.
\end{cases}
\end{equation}
We note, importantly, that in all the qubit-qutrit scenarios in 
which $\nu_{1}$ and $\nu_{2}$ have both appeared in the 
(naively, bivariate) separability
function, it has been in the
{\it product} form $\nu_1 \nu_2$ (cf. sec.~\ref{qubqut2}).
\section{Qutrit-Qutrit Analyses} \label{qutritqutrit}
In the qubit-qubit ($4 \times 4$ density matrix) case, we were able
to express the condition (\ref{Thirdcontion}) 
that the determinant of the partial transpose of
$\rho$ be nonnegative in terms of {\it one} supplementary variable ($\nu$), 
given by (\ref{BlooreRatio}),
rather than three independent diagonal entries.
Similarly, in the qubit-qutrit ($6 \times 6$ density matrix) case,
we could employ {\it two} supplementary variables ($\nu_{1}, \nu_{2}$), 
given by (\ref{tworatios}), rather 
than five independent diagonal entries.

For the qutrit-qutrit ($9 \times 9$ density matrix) case, rather than
eight independent diagonal entries, we found that one can employ
the {\it four} supplementary variables,
\begin{equation}
\nu_{1}= \frac{\rho_{11} \rho_{55}}{\rho_{22} \rho_{44}}; \hspace{.2in}
\nu_{2}= \frac{\rho_{22} \rho_{66}}{\rho_{33} \rho_{55}}; \hspace{.2in}
\nu_{3}= \frac{\rho_{44} \rho_{88}}{\rho_{55} \rho_{77}}; \hspace{.2in}
\nu_{4}= \frac{\rho_{55} \rho_{99}}{\rho_{66} \rho_{88}}.
\end{equation}
\subsection{Thirty-five nullified pairs of off-diagonal entries---36 
scenarios}
\subsubsection{10-dimensional complex case---$P^{HS}_{PPT} =\frac{1}{3}; \frac{1}{6}$}
Here, we nullify all but one of the thirty-six pairs of off-diagonal entries
of the $9 \times 9$ density matrix $\rho$. We allow this solitary pair to be
composed of complex conjugates. Since the Peres-Horodecki 
positive partial 
transposition (PPT) criterion is not sufficient to ensure separability,
we accordingly modify our notation.

Our first result is
\begin{equation}
\mathcal{S}_{[\tilde{(1,5)}]}^{9 \times 9}(\nu_{1}) = 
\begin{cases}
 \pi  & \nu _1\leq 1 \\
 \frac{\pi }{\nu _1} & \nu_{1} > 1
\end{cases}
\end{equation}
(a dual scenario being $[\tilde{(2,4)}]$).
We have $V^{HS}_{tot} =\frac{\pi}{3628800}, V^{HS}_{PPT}= 
\frac{\pi}{10886400}$, so $P^{HS}_{PPT}= \frac{1}{3}$.

The same three outcomes are obtained based on the
PPT function
\begin{equation}
\mathcal{S}_{[\tilde{(1,6)}]}^{9 \times 9}(\nu_{1},\nu_{2}) =
\begin{cases}
 \pi  & \nu_{1} \nu _2 \leq 1 \\
 \frac{\pi }{\nu _1 \nu _2} & \nu _1 \nu_{2} > 1 .
\end{cases}
\end{equation}
On the other hand, we have $V^{HS}_{tot} =\frac{\pi}{3628800}, V^{HS}_{PPT}=
\frac{\pi}{21772800}$, and  $P^{HS}_{PPT}= \frac{1}{6}$ based on the
PPT function
\begin{equation}
\mathcal{S}_{[\tilde{(6,8)}]}^{9 \times 9}(\nu_{4}) =
\begin{cases}
 \pi  & \nu _4\geq 1  \\
 \pi  \nu _4 & 0<\nu _4<1.
\end{cases}
\end{equation}

Of the thirty-six combinatorially possible scenarios, thirteen had
$P^{HS}_{PPT} = \frac{1}{3}$, while four had $P^{HS}_{PPT} = \frac{1}{3}$, 
and the remaining nineteen were fully separable in nature.
\subsection{Thirty-four  nullified pairs of off-diagonal entries---630 scenarios}
\subsubsection{12-dimensional complex case---$P^{HS}_{PPT} =\frac{1}{3}; \frac{7}{30}$}
Since the number of combinatorially possible scenarios was so large,
we randomly generated scenarios to examine.

Firstly, we found
\begin{equation}
\mathcal{S}_{[\tilde{(1,4)},\tilde{(3,5)}]}^{9 \times 9}(\nu_{2}) =
\begin{cases}
 \pi ^2 & \nu _2\geq 1 \\
 \pi ^2 \nu _2 & 0<\nu _2<1.
\end{cases}
\end{equation}
For this scenario, we had 
$V^{HS}_{tot}= \frac{\pi ^2}{479001600}, 
V^{HS}_{PPT} = \frac{\pi^2}{1437004800}$, 
giving us $P^{HS}_{PPT}= \frac{1}{3}$.

Also, we found
\begin{equation}
\mathcal{S}_{[\tilde{(2,9)},\tilde{(6,9)}]}^{9 \times 9}(\nu_{2},\nu_{4}) =
\begin{cases}
 \frac{\pi ^2}{2} & \nu _2  \nu_{4} \leq 1  \\
 \frac{\pi ^2 \left(2 \nu _2 \nu _4-1\right)}{2 \nu _2^2
   \nu _4^2} & \nu _2  \nu_{4} >1.
\end{cases}
\end{equation}
For this scenario, we had
$V^{HS}_{tot}= \frac{\pi ^2}{479001600},
V^{HS}_{PPT} = \frac{\pi^2}{2052864000}$,
giving us $P^{HS}_{PPT}= \frac{7}{30} \approx 0.23333$.
\section{Qubit-Qubit-Qubit Analyses, I} \label{qubitqubitqubitI}
For initial relative simplicity, let us 
regard an $8 \times 8$ density matrix $\rho$ as a {\it bipartite}
system, a composite of a four-level system and a two-level system.
Then, we can compute the partial transposition of $\rho$, transposing in
place its four $4 \times 4$ blocks. 
The nonnegativity of this partial transpose can be expressed using just
{\it three} ratio variables,
\begin{equation}
\nu_{1} = \frac{\rho_{11} \rho_{66}}{\rho_{22} \rho_{55}}; \hspace{.2in}
\nu_{2} = \frac{\rho_{22} \rho_{77}}{\rho_{33} \rho_{66}}; \hspace{.2in}
\nu_{3} = \frac{\rho_{33} \rho_{88}}{\rho_{44} \rho_{77}},
\end{equation}
rather than seven independent diagonal entries.
\subsection{Twenty-seven nullified pairs of off-diagonal entries---28
scenarios}
\subsubsection{9-dimensional complex case---$P^{HS}_{PPT}= \frac{1}{3}$}
We have the PPT function
\begin{equation}
\mathcal{S}_{[\tilde{(1,6)}]}^{8 \times 8}(\nu_{1}) =
\begin{cases}
 \pi  & \nu _1\leq 1 \\
 \frac{\pi }{\nu _1} & \nu_{1} > 1.
\end{cases}
\end{equation}
(Scenario $[\tilde{(2,5)}]$ was dual to this one.)
For this scenario, $V^{HS}_{tot}= \frac{\pi}{362880},
V^{HS}_{PPT}= \frac{\pi}{1088640}$, yielding
$P^{HS}_{PPT}= \frac{1}{3}$.
There were twelve scenarios, {\it in toto}, with 
precisely these three outcomes.
The other sixteen were all fully separable in nature.
\subsection{Twenty-six nullified pairs of off-diagonal entries---378
scenarios}
\subsubsection{11-dimensional complex case---$P^{HS}_{PPT} =\frac{1}{3}; \frac{1}{9}$}
Again, because of the large number of possible scenarios,
we chose them randomly for inspection.

Firstly, we obtained
\begin{equation}
\mathcal{S}_{[\tilde{(3,5)},\tilde{(6,8)}]}^{8 \times 8}(\nu_{1},\nu_{2}) =
\begin{cases}
 \pi ^2 & \nu _1>0\land \frac{1}{\nu _1}\leq \nu _2 \\
 \pi ^2 \nu _1 \nu _2 & \nu _1>0\land \frac{1}{\nu _1}>\nu
   _2\land \nu _2>0.
\end{cases}
\end{equation}
(Of course, the symbols ``$\land$'' and ``$\lor$'', used by Mathematica
in its output, denote the logical connectives ``and'' (conjunction) and ``or''
 (intersection) of propositions.)
For this scenario, we had $V^{HS}_{tot} =
\frac{\pi^2}{39916800}, V^{HS}_{PPT} = \frac{\pi^2}{119750400}$, 
giving us $P^{HS}_{PPT}=\frac{1}{3}$.

Also,
\begin{equation}
\mathcal{S}_{[\tilde{(2,5)},\tilde{(4,7)}]}^{8 \times 8}(\nu_{1},\nu_{2}) =
\begin{cases}
 \pi ^2 & \nu _1\geq 1\land \nu _3\geq 1 \\
 \pi ^2 \nu _1 & 0<\nu _1<1\land \nu _3\geq 1 \\
 \pi ^2 \nu _3 & \nu _1\geq 1\land 0<\nu _3<1 \\
 \pi ^2 \nu _1 \nu _3 & 0<\nu _1<1\land 0<\nu _3<1.
\end{cases}
\end{equation}
For this scenario, we had $V^{HS}_{tot} =
\frac{\pi^2}{39916800}, V^{HS}_{PPT} = \frac{\pi^2}{359251200}$,
giving us $P^{HS}_{PPT}=\frac{1}{9}$.

We also found the PPT function
\begin{equation}
\mathcal{S}_{[\tilde{(1,3)},\tilde{(4,7)}]}^{8 \times 8}(\nu_{1}) =
\begin{cases}
 \frac{\pi ^2}{2} & \nu _3=1 \\
 \frac{\pi ^2 \nu _3}{2} & 0<\nu _3<1 \\
 \pi  \left(\cos ^{-1}\left(\sqrt{1-\frac{1}{\nu
   _3}}\right)-\sin^{-1}\left(\frac{1}{\sqrt{\nu _3}}\right)\right)
   \nu _3+\pi ^2-\frac{\pi ^2}{2 \nu _3} & \nu _3>1
\end{cases}.
\end{equation}

\section{Qubit-Qubit-Qubit Analyses. II} \label{qubitqubitqubitII}
Here we regard the $8 \times 8$ density matrix as a {\it tripartite}
composite of three two-level systems, and compute the partial transpose by
transposing in place the eight $2 \times 2$ blocks of $\rho$.
(For {\it symmetric} states of three qubits, positivity of the partial
transpose is {\it sufficient} to ensure separability \cite{eckert,wangfeiwu}.)
Again the nonnegativity of the determinant could be expressed using
three (different) ratio variables,
\begin{equation}
\nu_{1} = \frac{\rho_{11} \rho_{44}}{\rho_{22} \rho_{33}}; \hspace{.2in}
\nu_{2} = \frac{\rho_{44} \rho_{55}}{\rho_{33} \rho_{66}}; \hspace{.2in}
\nu_{3} = \frac{\rho_{55} \rho_{88}}{\rho_{66} \rho_{77}},
\end{equation}
\subsection{Twenty-seven nullified pairs of off-diagonal entries---28
scenarios}
\subsubsection{9-dimensional complex case---$P^{HS}_{PPT}= \frac{1}{3}$}
There were, again, twelve of twenty-eight scenarios with non-trivial
separability properties, all with 
 $V^{HS}_{tot}= \frac{\pi}{362880},
V^{HS}_{PPT}= \frac{\pi}{1088640}$, yielding
$P^{HS}_{PPT}= \frac{1}{3}$. 
One of these was
\begin{equation}
\mathcal{S}_{[\tilde{(1,4)}]}^{8 \times 8}(\nu_{1}) =
\begin{cases}
 \pi  & \nu _1\leq 1 \\
 \frac{\pi }{\nu _1} & \nu_{1} > 1.
\end{cases}
\end{equation}
\subsubsection{11-dimensional complex case---$P^{HS}_{PPT}= \frac{17}{60}; \frac{1}{3}$}
We obtained the PPT function
\begin{equation}
\mathcal{S}_{[\tilde{(1,8)},\tilde{(5,7)}]}^{8 \times 8}(\nu_{1},\nu_{2},\nu_{3}) =
\begin{cases}
 \frac{\pi ^2}{2} & \frac{\nu _2}{\nu _1}=\nu _3\land \nu
   _1>0\land \nu _2>0 \\
 \pi ^2 & \nu _2>0\land \left(\left(\nu _1=0\land \nu
   _3\geq 0\right)\lor \left(\nu _3=0\land \nu
   _1>0\right)\right) \\
 \frac{\pi ^2 \nu _2}{4 \nu _1 \nu _3} & \nu _1>0\land \nu
   _2>0\land \frac{\nu _2}{\nu _1}<\nu _3 \\
 \pi ^2-\frac{\pi ^2 \nu _1 \nu _3}{2 \nu _2} & \nu
   _1>0\land \nu _2>0\land \frac{\nu _2}{\nu _1}>\nu
   _3\land \nu _3>0.
\end{cases}
\end{equation}
For this we had $V_{tot}^{HS}= \frac{\pi^2}{39916800},
V_{PPT}^{HS} =\frac{17 \pi^2}{239500800}$, giving us
$P^{HS}_{PPT}= \frac{17}{60} \approx 0.283333$.

Additionally,
\begin{equation}
\mathcal{S}_{[\tilde{(1,4)},\tilde{(7,8)}]}^{8 \times 8}(\nu_{1})= 
\begin{cases}
 \pi ^2 & \nu _1\leq 1 \\
 \frac{\pi ^2}{\nu _1} & \nu_{1} > 1.
\end{cases}
\end{equation}
Here, we had $V_{tot}^{HS}= \frac{\pi^2}{39916800},
V_{PPT}^{HS} =\frac{17 \pi^2}{119750400}$, giving us
$P^{HS}_{PPT}= \frac{1}{3}$.

Another PPT function we were able to find was
\begin{equation}
\mathcal{S}_{[\tilde{(3,4)},\tilde{(3,8)}]}^{8 \times 8}(\nu_{2},\nu_{3})=
\begin{cases}
 \frac{\pi ^2}{2} & \nu _2>0\land \left(\nu _3=\nu _2\lor
   \left(\nu _2>\nu _3\land \nu _2<2 \nu _3\right)\lor
   \left(\nu _2\geq 2 \nu _3\land \nu _3\geq
   0\right)\right) \\
 \frac{\pi ^2 \nu _2 \left(2 \nu _3-\nu _2\right)}{2 \nu
   _3^2} & \nu _2>0\land \nu _2<\nu _3.
\end{cases}
\end{equation}

\section{Approximate Approaches to 9-Dimensional Real Qubit-Qubit Scenario} \label{approximate}
As we have earlier noted, it appears that the 
simultaneous computational 
enforcement of the four conditions 
((\ref{firstcondition}), (\ref{secondcondition}), 
(\ref{2by2}), (\ref{Thirdcontion})) 
that would yield us the 9-dimensional
volume of the separable real two-qubit states appears presently 
highly intractable.
But if we replace (\ref{Thirdcontion})  by {\it less} strong conditions on the
nonnegativity of the partial transpose ($\rho^{T}$), 
we can achieve some form of
approximation to the desired results. So, replacing 
(\ref{Thirdcontion}) 
by the requirement (derived from a $2 \times 2$ principal minor 
of $\rho^{T}$) that 
\begin{equation} \label{minor1}
1 - \nu z_{14}^2 \geq 0,
\end{equation} 
we obtain the
{\it approximate} separability function
\begin{equation} \label{firstminorapprox}
S_{real}^{approx}(\nu)=
\begin{cases}
 \frac{512 \pi ^2}{27} & 0 <  \nu \leq 1 \\
 \frac{256 \left(3 \pi ^2 \nu -\pi ^2\right)}{27 \nu
   ^{3/2}} & \nu > 1.
\end{cases}
\end{equation}
(In the analyses in this section, we utilize the integration limits
on the $z_{ij}$'s \cite[eqs. (3)-(5)]{slaterPRA2} 
yielded by the cylinrical decomposition algorithm [CAD], to reduce the
dimensionalities of our constrained integrations.)
This yields an {\it upper bound} on the separability probability
of the real 9-dimensional qubit-qubit states of
$\frac{1}{2}+\frac{512}{135 \pi ^2} \approx 0.88427$. 
We obtain the same probability if we employ instead of (\ref{minor1}) the
requirement
\begin{equation} \label{minor2}
\nu - z_{23}^2 \geq 0,
\end{equation}
which yields the dual function to (\ref{firstminorapprox}), namely,
\begin{equation} \label{secondminorapprox}
S_{real}^{approx}(\nu)=
\begin{cases}
 \frac{512 \pi ^2}{27} & \nu \geq 1 \\
 \frac{256}{27} \pi ^2 (3-\nu) \sqrt{\nu } & 0<\nu <1.
\end{cases}
\end{equation}
(The left-hand sides of (\ref{minor1}) and (\ref{minor2}) are the only
two of the six $2 \times 2$ principal minors of $\rho^{T}$ that are
non-trivially distinct---apart from cancellable 
nonnegative factors---from the corresponding minors of $\rho$ itself.)
The non-constant functional form in the second line of 
(\ref{secondminorapprox}) will
emerge again, importantly, in (\ref{newbeta}).

If we form a ``quasi-separability'' function over 
$\nu \in [0,\infty]$ by piecing together
the non-constant segments of (\ref{firstminorapprox}) and 
(\ref{secondminorapprox}), we can
infer---using a simple symmetry, duality argument---an improved 
(lowered) upper 
bound on the HS separability probability of
$\frac{1024}{135 \pi^2} \approx 0.76854$.
We can also reach such a result by noting that the two constraints
(\ref{minor1}) and (\ref{minor2}) 
are independent (involve different variables), so we should
just be able to multiply the corresponding functions (and then scale them
by the corresponding $c_{scenario} = \frac{512 \pi^2}{27}$)
\section{Alternative use of Bloch parameterization}
We may say, in partial summary that we have been able to obtain 
certain {\it exact}
two-qubit HS separability probabilities in dimensions 
seven or less, making use of the 
advantageous Bloore parameterization \cite{bloore}, but not yet in
dimensions greater than seven. This, however, is considerably greater than 
simply the three dimensions (parameters)
we were able to achieve \cite{pbsJak} in a somewhat comparable study based on 
the {\it generalized Bloch representation} parameterization \cite{kk,jak}.
In \cite{pbsJak}---extending an approach of Jak\'obczyk and 
Siennicki \cite{jak}---we 
primarily studied {\it two}-dimensional sections of a set of 
generalized Bloch vectors corresponding to $n \times n$ density matrices, 
for $n=4, 6, 8, 9$ and 10. For $n>4$, by far the most frequently recorded
HS separability [or positive partial transpose (PPT) for $n>6$]
probability was $\frac{\pi}{4} \approx 0.785398$. A very wide range
of exact HS separability and PPT probabilities were tabulated.

Immediately below is just one of many matrix tables 
(this one being numbered (5)) presented in \cite{pbsJak} 
(which due to its copious results has been left 
simply as a preprint, rather than 
submitted directly to a journal).
This table gives the HS separability probabilities for the
qubit-qutrit case. In the first column are given the identifying numbers
of a {\it pair} of generalized Gell-mann matrices (generators of $SU(6)$).
In the second column of (\ref{n=6case1}) 
are shown the {\it number} of distinct unordered pairs
of $SU(6)$ generators which share the same total (separable and nonseparable)
HS volume, as well as the same separable HS volume, and consequently,
identical HS separability probabilities. The third column gives us these
HS total volumes, the fourth column, the HS separability probabilities and 
the last (fifth) column, numerical approximations to the exact probabilities 
(which, of course, we see---being probabilities---do not exceed the  value 1).
 (The HS separable volumes too can 
be deduced from the total volume and the separability probability.)

\begin{equation} \label{n=6case1}
\left(
\begin{array}{lllll}
 \{1,13\} & 48 & \frac{4}{9} & \frac{\pi }{4} & 0.785398
   \\
 \{3,11\} & 4 & \frac{8 \sqrt{2}}{27} & \frac{1}{\sqrt{2}}
   & 0.707107 \\
 \{3,13\} & 4 & \frac{4}{9} & \frac{5}{6} & 0.833333 \\
 \{3,25\} & 4 & \frac{8 \sqrt{2}}{27} & \frac{5}{4
   \sqrt{2}} & 0.883883 \\
 \{8,13\} & 4 & \frac{2}{3} & \frac{1}{\sqrt{3}} &
   0.577350 \\
 \{8,25\} & 4 & \frac{\sqrt{2}}{3} & \sqrt{\frac{2}{3}} &
   0.816497 \\
 \{11,15\} & 4 & \frac{4 \sqrt{2} \pi }{27} &
   \frac{1}{3}+\frac{3 \sqrt{3}}{4 \pi } & 0.746830 \\
 \{11,24\} & 2 & \frac{25 \sqrt{\frac{5}{2}}}{72} &
   \frac{2}{5}+\frac{1}{2} \sin
   ^{-1}\left(\frac{4}{5}\right) & 0.863648 \\
 \{13,24\} & 2 & \frac{25 \sqrt{\frac{5}{2}}}{72} &
   \frac{8}{75} \left(-2+5 \sqrt{5}\right) & 0.979236 \\
 \{13,35\} & 4 & \frac{4 \sqrt{\frac{3}{5}}}{5} &
\frac{1}{12} \left(5+3 \sqrt{5} \sin^{-1}
   \left(\frac{\sqrt{5}}{3}\right)\right) & 0.886838
   \\
 \{15,16\} & 4 & \frac{32 \sqrt{2}}{81} & \frac{1}{32}
   \left(9 \sqrt{3}+4 \pi \right) & 0.879838 \\
 \{16,24\} & 2 & \frac{25}{144} \sqrt{\frac{5}{2}} \pi  &
   \frac{4+5 \sin ^{-1}\left(\frac{4}{5}\right)}{5 \pi } &
   0.549815 \\
 \{20,24\} & 2 & \frac{25}{144} \sqrt{\frac{5}{2}} \pi  &
   \frac{92+75 \sin ^{-1}\left(\frac{4}{5}\right)}{75 \pi
   } & 0.685627 \\
 \{24,25\} & 2 & \frac{25}{27 \sqrt{2}} & 1-\frac{2}{5
   \sqrt{5}} & 0.821115 \\
 \{24,27\} & 2 & \frac{25}{27 \sqrt{2}} & \frac{92+75 \cos
   ^{-1}\left(\frac{3}{5}\right)}{80 \sqrt{5}} & 0.903076
   \\
 \{25,35\} & 4 & \frac{\sqrt{3} \pi }{5} &
   \frac{\sqrt{5}+3 \csc
   ^{-1}\left(\frac{3}{\sqrt{5}}\right)}{3 \pi } &
   0.504975
\end{array}
\right).
\end{equation}
It might be of interest to address separability problems that appear to be
computationally intractable in the generalized Bloch representation by
transforming them into the Bloore parameterization.
\section{{\it Full} real and complex two-qubit 
separability probability conjectures} \label{twoqubitconjectures}
The qubit-qubit results above (sec.~\ref{qubit-qubit}) 
motivated us to reexamine 
previously obtained results
(cf. \cite[eqs. (12), (13)]{univariate}) and we would like to 
make the following observations pertaining to
the {\it full} 9-dimensional real and 15-dimensional HS separability 
probability issue.
We have the exact results in these two cases that
\begin{equation} \label{jacrealInt}
\int_{0}^{\infty} \mathcal{J}_{real}(\nu) = 2 \int_{0}^{1} \mathcal{J}_{real}(\nu) =\frac{\pi^2}{1146880} \approx 8.60561 \cdot 10^{-6}
\end{equation}
and
\begin{equation} \label{jaccomplexInt}
\int_{0}^{\infty} \mathcal{J}_{complex}(\nu) = 2 \int_{0}^{1} \mathcal{J}_{complex}(\nu) =\frac{1}{1009008000} \approx 
9.91072 \cdot 10^{-10} .
\end{equation}
Now, to obtain the corresponding {\it total} (separable plus nonseparable) 
 HS volumes computed by \.Zyczkowski and Sommers \cite{szHS}, that is,
$\frac{\pi^4}{60480} \approx 0.0016106$ 
and $\frac{\pi^6}{851350500} \approx 1.12925 \cdot 10^{-6}$, one must multiply
(\ref{jacrealInt}) and (\ref{jaccomplexInt}) by the 
factors of $C_{real}= \frac{512 \pi^2}{27} = \frac{2^8 \pi^2}{3^3} \approx  
187.157$ and 
$C_{complex} = 
\frac{32 \pi^6}{27} = \frac{2^5 \pi^6}{3^3} \approx 1139.42$, respectively.

To most effectively compare these previously-reported results with 
those derived above in this paper, one needs to multiply 
 $C_{real}$ and $\mathcal{S}_{real}(\nu)$,
by $2^{-4}=\frac{1}{16}$ and in the complex case by
$2^{-7}=\frac{1}{128}$. Doing so, for example, 
would adjust $C_{real}$ to equal
$c_{real} = \frac{32 \pi^2}{27} \approx 11.6973$, which we note, in line with
our previous series of calculations [sec.~\ref{sec8dim}] 
is equal to $\frac{4}{3} (\frac{8 \pi^2}{9}$). (Andai \cite{andai} 
also computed the same
volumes---up to a normalization factor---as 
\.Zyczkowski and Sommers \cite{szHS}.)
Now, our estimates from \cite{slaterPRA2} are that 
$\mathcal{S}_{real}(1)= 114.62351 < C_{real}$ and
$\mathcal{S}_{complex}(1) = 387.50809 < C_{complex}$. 
These results would appear---as remarked above---to 
be a reflection of the phenomena that there
are non-separable states for both the 9- and 15-dimensional scenarios
at $\nu=1$ (the locus of the fully mixed, classical state).

Alternatively, 
the results in sec.~\ref{Dyson}, and further throughout the paper, 
in which we find a relation 
between separability functions and the Dyson indices ($\beta=1, 2, 4$) of 
random matrix theory---including the frequent occurrence of $\sqrt{\nu}$ in
a real scenario and $\nu$ in the corresponding complex scenario---strongly 
suggest that in the full ($m=0$) 
9-dimensional real
and 15-dimensional complex cases scenarios, 
the separability function for the complex
case might simply be proportional to the {\it square} of the separability
function for the real case (and, in the quaternionic case \cite{adler}, 
to the {\it fourth}  power of that function). 

\begin{figure}
\includegraphics{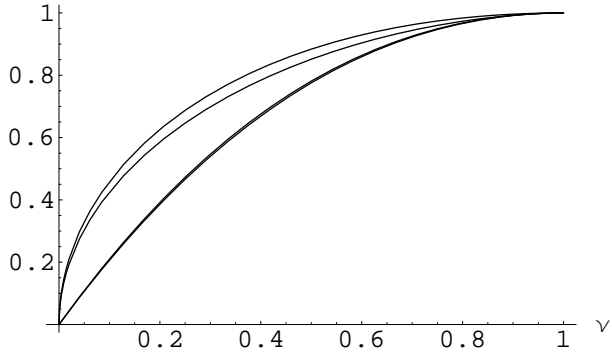}
\caption{\label{fig:scaled}The most subordinate of the three curves is actually
two virtually indistinguishable curves: (1) our  normalized 
(previously obtained \cite{slaterPRA2})
numerical estimate of $\mathcal{S}_{complex}(\nu)$;  
and (2) the (extraordinarily  well-fitting) {\it square} of the incomplete
beta function ratio $I_{\nu}(\nu,\frac{1}{2},2) = 
\frac{1}{2} (3-\nu) \sqrt{\nu }$. The intermediate
curve is the normalized previously obtained 
numerical estimate of $\mathcal{S}_{real}(\nu)$ and
the rather well-fitting (most  dominant)  
curve is $I_{\nu}(\nu,\frac{1}{2},2)$.}
\end{figure}

Following such a line of thought, we were led to reexamine 
the numerical analyses
reported in \cite{slaterPRA2}, in which we had formulated our beta function
ans{\"a}tze.
In Fig.~\ref{fig:scaled} we show the previously-obtained 
numerical estimates of $\mathcal{S}_{real}(\nu)$ and 
$\mathcal{S}_{complex}(\nu)$, now both scaled 
(``regularized'') 
to equal 1 at $\nu=1$, along with the 
similarly regularized form (termed the ``incomplete beta function 
ratio'' \cite{handbook} or, alternatively, the ``regularized 
incomplete beta function'')
\begin{equation} \label{newbeta}
I_{\nu}(\nu,\frac{1}{2},2) = \frac{1}{2} (3 -\nu) \sqrt{\nu}
\end{equation}
of the incomplete beta 
function,
$B_{\nu}(\nu,\frac{1}{2},2) = \frac{2}{3} (3 -\nu) \sqrt{\nu}$)
and $I_{\nu}(\nu,\frac{1}{2},2)^2$. (Let us make the important observation
here that the functional form (\ref{newbeta}) has---up to proportionality---already occurred [although we did not 
immediately perceive then its beta function expression] 
in certain  previous exact qubit-qubit analyses 
(\ref{secondminorapprox}) (see also (\ref{puzzle}) 
(\ref{morepuzzle}), but also (\ref{puzzle2}), for its occurrence in the 
qubit-qutrit context).

Fig.~\ref{fig:scaled} does reveal an extraordinarily good fit between the 
normalized numerical
estimates of $\mathcal{S}_{complex}(\nu)$ and $I_{\nu}(\nu,\frac{1}{2},2)^2$, 
while 
$I_{\nu}(\nu,\frac{1}{2},2)$
itself provides a close fit to the normalized numerical
estimates of $\mathcal{S}_{real}(\nu)$. (Note that 
$I_{\nu}(\nu,\frac{1}{2},2)$ does 
contain a factor of $\sqrt{\nu}$ and 
$I_{\nu}(\nu,\frac{1}{2},2)^2$, obviously a factor
of $\nu$, much in line with the more elementary lower-dimensional 
real-complex examples
studied in sec.~\ref{qubit-qubit}.
Further, as an exercise, we sought that value of $x$ for which 
the function $I_{\nu}(\nu,\frac{1}{2},x)$ would when employed in our basic 
paradigm here, as in Fig.~\ref{fig:scaled},
{\it jointly} minimize the sum of a certain least-squares
fit to the normalized numerical estimates of
$S_{real}(\nu)$ and $S_{complex}(\nu)$. Our Mathematica program produced the
answer $x=1.88487$, being somewhat intermediate in value between $\sqrt{3} 
\approx 1.732$ and 2, the exact candidate 
values we have considered, which both fit
the numerical results of \cite{slaterPRA2} rather well.)
So, it would seem appropriate to revise the two central ans{\"a}tze 
put forth in \cite{slaterPRA2} to account for these 
interesting newly-observed phenomena inherent in the results 
already reported in 
\cite{slaterPRA2}.

We can now exactly perform the 
requisite integrations (cf. (\ref{Vsmall}), (\ref{Vbig})),
\begin{equation} \label{almost1}
2 \int_{0}^{1} \mathcal{J}_{real}(\nu)  I_{\nu}(\nu,\frac{1}{2},2) d \nu = 
\frac{1}{151200} = \frac{1}{2^5 \cdot 3^3 \cdot 5^2 \cdot 7}, 
\end{equation}
\begin{equation} \label{almost2}
2 \int_{0}^{1} \mathcal{J}_{complex}(\nu)  I_{\nu}(\nu,\frac{1}{2},2)^2 d \nu =
\frac{71}{99891792000} = 
\frac{71}{2^7 \cdot 3^4 \cdot 5^3 \cdot 7^2 \cdot 11^2 \cdot 13},
\end{equation}
\begin{equation} \label{almost3}
2 \int_{0}^{1} \mathcal{J}_{quaternionic}(\nu)  I_{\nu}(\nu,\frac{1}{2},2)^4 d \nu =\frac{5989}{358347086242825680000}
\end{equation}
\begin{displaymath}
\frac{53 \cdot 113}{2^{7} \cdot 3^4 \cdot 5^4 \cdot 7^2 \cdot 11^2 \cdot 13^2 
\cdot 17^2 \cdot 19^2 \cdot 23^2}.
\end{displaymath}
The three marginal univariate jacobian functions above 
are obtained by transforming
the jacobian for the Bloore parameterization---($\Pi_{i=1}^{4} 
\rho_{ii})^{\frac{3 \beta}{2}}$, $\beta =1, 2, 4$---to the $\nu$ 
variable and integrating over the two remaining 
independent diagonal entries of $\rho$.

So, assuming the validity of our modified 
beta function ans{\"a}tze for the real and complex 
separability functions, all we still lack
for obtaining the Hilbert-Schmidt separable volumes/probabilities
of the 9-dimensional real and 15-dimensional complex qubit-qubit systems 
themselves 
are the appropriate (presumptively, exact in nature) 
scaling constants (on the order of 114.61 and 
387.467 \cite{slaterPRA2}) by which to multiply the results of 
(\ref{almost1}) and (\ref{almost2}). 
(We will presume---in light of the numerous analyses reported earlier---
that such scaling constants are exact in nature, being of the form
$\frac{i \pi^k}{j}$, where $i,j,k$ are natural numbers. We search over
the spaces of possibilities to find choices that accord with our
previously-obtained numerical results for the HS separability probabilities.)
\subsection{Real Two-Qubit Case}
If we employ $\frac{20 \pi^4}{17} \approx 114.599$ as the scaling
constant in the real case---giving us a very good fit to the numerical 
estimate of $\mathcal{S}_{real}(1) \approx 
114.61$---we obtain an HS separable volume of
$\frac{\pi^4}{128520}$ and an HS separability probability of $\frac{8}{17} 
\approx 0.470588$. (Using the numerical results of \cite{slaterPRA2}, we were 
able to obtain an estimate of this probability as close as 0.46968 by
replacing the
jacobian function (\ref{Jacreal}) 
by a sixth-order Taylor series approximation of it around 
$\nu= \frac{13}{16}$. Providing inferior fits to 114.61, but still 
of possible interest, would be choices of scaling constants
$\frac{7 \pi^4}{6} \approx 113.644$ and $\frac{32 \pi^4}{27} 
\approx 115.448$. These would lead to HS real separability probabilities
of $\frac{7}{15} \approx 0.466667$ and $\frac{64}{135} \approx 0.474074$---with the first of these two seeming much more consistent with the numerics 
of \cite{slaterPRA2} than the second.)

 By the (``twofold'') 
theorem of Szarek, Bengtsson and \.Zyczkowski 
\cite{sbz} (cf. \cite{innami})---formalizing results in 
\cite{slaterPRA}---the HS separable volume of the generically rank-3 real
qubit-qubit states
would---adopting $\frac{20 \pi^4}{17}$ as the appropriate scaling 
constant---be $\frac{\pi^4}{4760 \sqrt{3}}$ and the HS 
separability probability, $\frac{4}{17} \approx 0.235294$. 
(The HS area-volume ratio for the 9-dimensional real two-qubit states is
$18 \sqrt{3} \approx 31.1769$ \cite[eq. (7.9)]{szHS}, 
while the analogous ratio restricted
to the separable subset is one-half as large, that is, 
$9 \sqrt{3} \approx 15.5885$, indicating
the more hyperspherical-like shape of the separable subset).
\subsection{Complex Two-Qubit Case}
One simple candidate for the scaling  constant in the complex case is
$\frac{2 \pi^6}{5} \approx 384.566$. 
This   would yield an HS separability probability of  $\frac{213}{880} 
\approx 0.242045$.
But considerably 
more attractive  (certainly, in part, due to its 
interesting consonance 
with the real results just advanced, and also the presence 
of $256 =2^{8}$ and $639 = 9 \cdot 71$, with the 71 in 
(\ref{almost2}), thus, being cancelled), it seems 
is $\frac{256 \pi^6}{639} \approx 
385.157$ (slightly closer also to our estimate of 387.467 from 
\cite{slaterPRA2}). This choice would  yield
an HS separable volume of $\frac{2 \pi^6}{7023641625} 
\approx 2.73758 \cdot 10^{-7}$, and separability probability of
$\frac{8}{33} \approx 0.242424$, {\it very} close to our 
previous numerically-derived estimate of 0.242575 
(implicitly given in \cite[between eqs. (41) and (42)]{slaterPRA}) 
(and only slightly
more than one-half of $\frac{8}{17}$).
(Let us also indicate that the 
HS area-volume ratio for the 15-dimensional complex two-qubit states is
$30 \sqrt{3} \approx 51.9615$ \cite[eq. (6.5)]{szHS},
while the analogous ratio restricted
to the separable subset is again one-half, that is 
$15 \sqrt{3} \approx 25.9808$, the lesser 
value indicating 
the more hyperspherical-like shape of the separable subset).

In the real and complex analyses just conducted, we have tacitly assumed---as we will also do in the succeeding, remaining ones---that the appropriate
scaling constants should be of the form $\frac{i \pi^k}{j}$, where, in 
addition to $i$ and $j$ being natural numbers, $k$
is identical to the power of $\pi$ occurring in the 
{\.Z}yczkowski-Sommers/Andai formulas for the corresponding total volumes. Doing so, at least seems 
plausible, in light of our numerous lower-dimensional analyses above.

The simplicity of two-qubit complex and real HS separability probabilities, 
$\frac{8}{33}$ and $\frac{8}{17}$, apparently stemming from the use of the
 modified beta function ans{\"a}tze, now leads us to examine if we can
generate  somewhat parallel HS separability 
probabilities for the 15-  and  35-dimensional real and 
complex qubit-{\it qutrit} cases.
\section{{\it Full} real and complex qubit-qutrit
separability probability conjectures}
\subsection{Real Qubit-Qutrit Case} \label{qubqut2}
\begin{figure}
\includegraphics{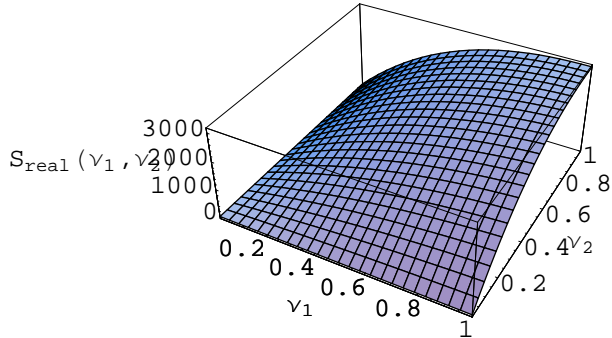}
\caption{\label{fig:QubQut}Interpolated estimate 
over the unit square of the real qubit-qutrit
separability function $S_{real}(\nu_{1},\nu_{2})$, based on 785,000
randomly generated $6 \times 6$ real density matrices}
\end{figure}

In Fig.~\ref{fig:QubQut} we show an 
interpolated estimate (with parameter values restricted to the unit square) 
of the {\it real} qubit-qutrit separability
function. (Auxiliary analyses give very strong evidence, as certainly 
seems plausible,  that this function  is 
{\it symmetric} under the interchange of $\nu_1$ and $\nu_2$.)
An immediate conjecture, suggested by our various earlier qubit-qutrit
results (sec.~\ref{qubitqutrit} and (\ref{puzzle})) 
is that this (naively, bivariate) function is actually univariate
in nature, {\it and} satisfies the proportionality relation
(cf. (\ref{newbeta}))
\begin{equation} \label{appealing}
\mathcal{S}_{real}(\nu_1,\nu_2) = \mathcal{S}_{real}(\eta) \propto 
I_{\eta}(\eta,\frac{1}{2},2) = \frac{1}{2} (3 -\eta) \sqrt{\eta}.
\end{equation}
Here $\eta=\nu_1 \nu_2 = \frac{\rho_{11} \rho_{66}}{\rho_{33} \rho_{44}}$, 
being {\it independent} of $\rho_{22}$  and $\rho_{55}$ (given the
definitions of $\nu_1$ and $\nu_2$ in eq. (\ref{tworatios}) above).
(If we had chosen to compute the partial transpose of $\rho$ by transposing
in place its nine $2 \times 2$ blocks, rather than its four $3 \times 3$ 
blocks, then presumably the same essential phenomenon would have occurred,
but with different sets of indices on the $\nu$'s.)
But, in fact, analyses we have conducted
indicate that it is the (even simpler) univariate
function satisfying the relation 
\begin{equation} \label{root}
\mathcal{S}_{real}(\nu_1,\nu_2) = \mathcal{S}_{real}(\eta) \propto 
\sqrt{\eta},
\end{equation}
that fits the sample estimate of the separability 
function displayed in Fig.~\ref{fig:QubQut} extremely well.

\begin{figure}
\includegraphics{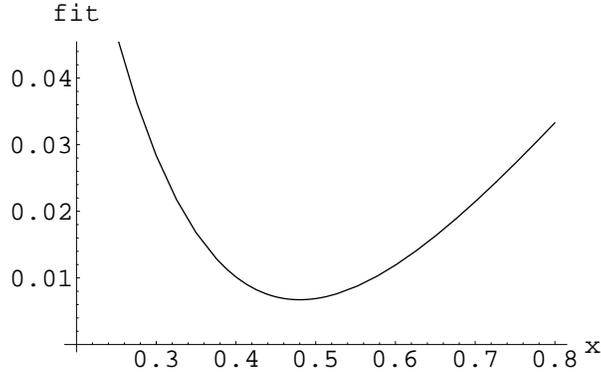}
\caption{\label{fig:prettygood}Least-squares fit of the 
normalized sample estimate of
the real qubit-qutrit separability function displayed in Fig.~\ref{fig:QubQut}
{\it vs.} $\eta^x = (\nu_1 \nu_2)^x$. The minimum of the curve is in
the immediate vicinity of $x=\frac{1}{2}$, at which point the measure of
goodness-of-fit is 0.00690362.}
\end{figure}

To substantiate this last 
point, in Fig.~\ref{fig:prettygood}, we show a plot of a least-squares 
fit of the normalized function
shown in Fig.~\ref{fig:QubQut} to the function $\eta^x$, which for $x = 
\frac{1}{2}$ is identical to (\ref{root}). 
We see that the best fit 
does, in fact, suggest that $x =\frac{1}{2}$ is the appropriate choice 
(at least, within this one-parameter family of functions). (For the 
least-squares fit of
(\ref{root}) to our sample estimate, we obtain 0.00690362, 
while we obtain considerably more, that is 0.0184813, for the [inferior]
fit of (\ref{appealing}).)
The product of the normalized function 
(\ref{root}) with the corresponding
jacobian is integrable in the real qubit-qutrit case 
(giving the result $\frac{131 \pi }{1110124175582822400} \approx 
3.70723 \cdot 10^{-16}$).

If we adopt the ansatz (\ref{root}) and employ the estimated value of
the scaling constant for this function from Fig.~\ref{fig:QubQut}, which is
on the order of 3095.97, {\it and} additionally presume that the real
qubit-qutrit HS separability probability (in line with our complex
counterpart conjecture [immediately below]  
of $\frac{32}{1199}$---and 
qubit-qubit proposals of $\frac{8}{33}$ and 
$\frac{8}{17}$) is of the form $\frac{32}{k}$, 
where $k$ is some natural number, then our best estimate of this
probability is $\frac{32}{213} \approx 0.150235$, and of 
the scaling constant $\frac{78848 \pi ^8}{139515 \sqrt{3}}
\approx 3096.05$. (We do not have
highly extensive numerical estimates [only the more limited one pursued 
here] ---as we did in the complex qubit-qutrit 
case \cite{slaterPRA}---against which to gauge this prediction, but our
fits here are strongly supportive of these assertions. For example, our
{\it sample} estimate of the separability probability can be expressed as
$\frac{32}{213.005}$.)
\subsection{Complex Qubit-Qutrit Case}
Based on our previous numerically-intensive study 
---using $10^9$ sample points ---- we have an (implicitly-given) 
estimate \cite[between eqs. (38) and (39)]{slaterPRA} 
for the complex HS separability probability of
0.0266891. 
A {\it very} well-fitting candidate for the corresponding exact
probability is $\frac{32}{1199} \approx 0.0266889$. 
(The associated separable volume would, then, be
$\frac{\pi ^{15}}{56980588975590080071885989375000
   \sqrt{6}} \approx 2.05327 \cdot 10^{-25}$.) Aside from the striking
goodness-of-fit, we see that the numerator of the probability 
is equal to
$32 = 2^{n-1}, n=6$, while in the qubit-qubit case, 
the numerator is  
$8 = 2^{n-1}, n=4$. Also, the denominator $1199 = 109 \cdot 11$, 
while $33 = 3 \cdot 11$.

In line with the Dyson-indices pattern observed earlier, we investigated
the possibility that 
the separability function in the complex qubit-qutrit case might be simply
proportional to $\eta= \nu_1 \nu_2$, that is, the square of 
its putative real counterpart, $\sqrt{\eta}$.
The integral of the  product
of $\eta$  with the associated jacobian 
yields  $\frac{829}{5045434342262725360252343040000} 
\approx 1.64307 \cdot 10^{-28}$.
With
our proposal above (supported by the considerable numerical evidence 
of \cite{slaterPRA2})
that the qubit-qutrit complex HS separability probability
is $\frac{32}{1199}$, the scaling constant would be
$\frac{537472 \sqrt{\frac{2}{3}} \pi ^{15}}{10063956375} \approx 
1249.65$.

\begin{figure}
\includegraphics{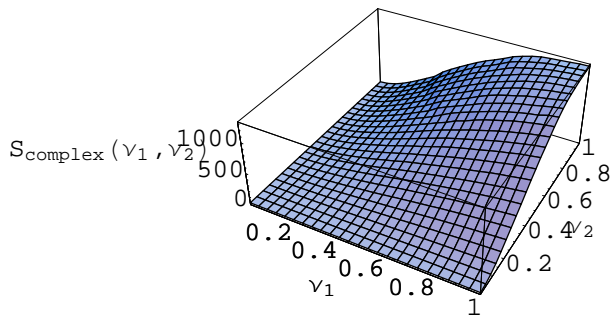}
\caption{\label{fig:QubQut2}Interpolated estimate
over the unit square of the complex qubit-qutrit
separability function $S_{complex}(\nu_{1},\nu_{2})$, based on 880,000
randomly generated $6 \times 6$ complex density matrices}
\end{figure}

\begin{figure}
\includegraphics{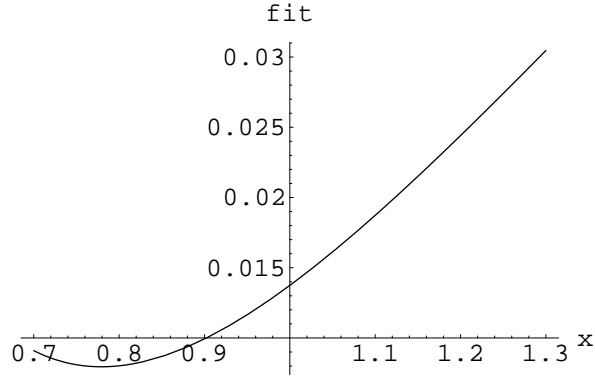}
\caption{\label{fig:prettygood2}Least-squares fit of the 
normalized sample estimate of
the complex 
qubit-qutrit separability function displayed in Fig.~\ref{fig:QubQut2}
{\it vs.} $\eta^x = (\nu_1 \nu_2)^x$. We hypothesize that for a 
sufficiently large sample the minimum would lie at $x=1$.}
\end{figure}

Figs.~\ref{fig:QubQut2} and \ref{fig:prettygood2} are the {\it 
complex} qubit-qutrit counterparts of the 
(real qubit-qutrit) Figs.~\ref{fig:QubQut} and \ref{fig:prettygood}.
Fig.~\ref{fig:prettygood2} might be said to weakly
support the proposal that the separability function is
 proportional to $\eta$. (The numerics here are perhaps 
yet insufficient
for our purposes. In addition to only so far having
sampled a relatively small
number of complex $6 \times 6$ density matrices, the sample
points are now 30-dimensional in nature. For alacrity, we had simply used
Monte Carlo methods, and not the [better-behaved/''lower-discrepancy''] 
quasi-Monte Carlo [Tezuka-Faure] methods employed in our
earlier studies, in particular, in \cite{slaterPRA,slaterPRA2}. 
In light of the not very convincing nature of Fig.~\ref{fig:prettygood2},
it might be advisable to revert to the Tezuka-Faure scheme, although most
of the unit hypercube points generated would, then, 
 be simply discarded as not meeting the criteria 
a density matrix must fulfill. The most desirable/efficient sampling 
scheme, it seems, if it can be effectively implemented, would be the one
associated with correlation matrices \cite{joe,kurowicka,kurowicka2,kurowicka3}, in which {\it none} of the generated points would have to be discarded.
\section{Concluding Remarks}
In a recent comprehensive  review, it was stated
 that while quantum entanglement 
is ``usually fragile to environment,
it is robust against conceptual and mathematical tools, the task of
which is to decipher its rich structure''  \cite[abstract]{family}.
We have attempted to make some progress in this 
regard here, but considerable impediments clearly 
still remain to putting
the chief conjectures of this paper on a fully rigorous basis 
(or disproving them by establishing alternative results), and 
in proceeding onward to higher-dimensional cases. 
(In particular, we have not yet developed a theory to predict 
the scaling constants---$\frac{256 \pi^6}{639}$ 
and $\frac{20 \pi^4}{17}$ in the full complex and real two-qubit cases, 
and $\frac{537472 \sqrt{\frac{2}{3}} \pi ^{15}}{10063956375}$ 
and $\frac{78848 \pi ^8}{139515 \sqrt{3}}$ in the full complex 
and real qubit-qutrit cases---for the hypothesized separability functions.)
It would seem that applications and/or extensions of 
random matrix theory and, possibly,  mathematical induction 
will be important---as they  were
in determining the total (separable {\it and} nonseparable) 
Hilbert-Schmidt volumes \cite{szHS} 
\cite[sec. 14.3]{ingemarkarol} \cite{andai}.

Let us, still, further suggest that the analytical framework and 
results, both theoretical and numerical,
presented above may lead to the development of associated formal 
propositions, much in the way that the numerically-obtained (two-fold)
separability-probability ratios reported in \cite{slaterPRA} led
Szarek, Bengtsson and {\.Z}ycskowski to establish that the 
set of separable (and, more generally, positive-partial-transpose)
states is a convex body of constant height \cite{sbz}.

In \cite{BuresSep}, we have 
further applied the ``separability
function'' concept to the determination of the {\it Bures} (minimal
monotone) metric volume of certain low-dimensional (real, complex and 
quaternionic) two-qubit states.
Interestingly, we find that although the Dyson-index pattern is not now
fully adhered too, it does come remarkably close to holding.
Also, numerical research that we hope to 
shortly report, strongly indicates 
that in the
full two-qubit {\it quaternionic} 
27-dimensional {\it Hilbert-Schmidt} separable volume case,
the Dyson-index pattern ($\beta =1, 2, 4$) we have observed above in the 
two-qubit 9-dimensional real 
and 15-dimensional complex cases (FIg.~\ref{fig:scaled}) is
strictly maintained.

\begin{acknowledgments}
I would like to express gratitude to the Kavli Institute for Theoretical
Physics (KITP)
for computational support in this research
and to a referee for indicating
that the apparent oscillatory nature of certain marginal jacobian functions
can be shown to be illusory if sufficiently high-precision is used in plotting.

\end{acknowledgments}

\bibliography{ReviseHS5}

\end{document}